\documentclass[final,5p,times,twocolumn]{elsarticle}

\usepackage{amssymb}
\usepackage{amsmath}
\usepackage{graphicx}
\usepackage{subcaption}

\usepackage{eurosym}

\usepackage{url}
\usepackage{bm}
\usepackage[dvipsnames]{xcolor}

\usepackage[normalem]{ulem}

\usepackage{nomencl}
\makenomenclature

\usepackage{etoolbox}
\renewcommand\nomgroup[1]{%
  \item[
  \ifstrequal{#1}{I}{Indices}{%
  \ifstrequal{#1}{V}{Variables}{
  \ifstrequal{#1}{P}{Parameters}{}}}%
]}

\begin{document}

\begin{frontmatter}



\title{Control and Scheduling of Behind-the-Meter Battery Energy Storage Systems for Stacked Grid and Building Services} %% Article title


\author{Nour-eddine Id omar, Alexandre Lê-Agopyan, Fabrizio Sossan} %% Author name

\affiliation{organization={HES-SO Valais - Wallis},%Department and Organization
            addressline={Rue de l'industrie 23}, 
            city={Sion},
            postcode={1950}, 
            state={},
            country={Switzerland}}

%% Abstract
\begin{abstract}
This paper proposes and experimentally validates a two-stage scheduling and control strategy for a behind-the-meter battery energy storage system (BESS) delivering both local and grid services. Considered services are the maximization of PV self-consumption, peak-load reduction, and secondary frequency control (aFRR).The day-ahead stage allocates battery capacity across local and balancing services using a scenario based approach, reflecting potential remuneration from aFRR participation without committing to fixed power availability; in the real-time stage, BESS set-points are computed in a periodic fashion at a high time resolution based on updated information on balancing prices, net load realization and BESS state of charge.
The strategy is experimentally validated on a building at the Energypolis Campus of HES-SO Valais (Sion, Switzerland), which exhibits a peak power demand of 300 kW and is equipped with a 264 kWh / 140 kW lithium-ion BESS. The experimental results demonstrate the effectiveness of the proposed framework in scheduling and actuating the provision of both behind-the-meter and front-of-the-meter services.
\end{abstract}

% \includegraphics[width=1\linewidth]{fig/Graphical abstract.drawio.png}

% Highglights
% Framework for providing local and grid services with behind-the-meter flexibility
% Integrated scheduling, real-time control, and forecasting schemes
% Tractable formulations implementable with off-the-shelf optimization software
% Experimental validation using a building BESS delivering local services and aFRR

%% Keywords

\end{frontmatter}

\begin{thenomenclature} 
\nomgroup{I}
  \item [{\(\omega\)}]\begingroup Scenario index , \(\omega \in \{0, 1, 2, \dots, \Omega\}\)\nomeqref {0}\nompageref{1}
  \item [{\(i\)}]\begingroup Day index, \(i \in \{0, 1, 2, \dots, N\}\)\nomeqref {0}\nompageref{1}
  \item [{\(k\)}]\begingroup Discrete time index $k \in \{0, 1, 2, \dots, K-1\}$ with 30-sec resolution.\nomeqref {0}\nompageref{1}
  \item [{\(k^*\)}]\begingroup Index for the current time interval\nomeqref {0}\nompageref{1}
  \item [{\(t\)}]\begingroup Discrete time index $t \in \{0,\dots,T-1\}$ with 15-min resolution.\nomeqref {0}\nompageref{1}
\nomgroup{P}
  \item [{\(\bm{\hat{L}_{Gross}^{i}}\)}]\begingroup Historical gross load profile for day i (kW)\nomeqref {0}\nompageref{1}
  \item [{\(\bm{\hat{L}_{Gross}}\)}]\begingroup Forecast of the building gross electrical load (kW)\nomeqref {0}\nompageref{1}
  \item [{\(\bm{\hat{P}_{PV}}\)}]\begingroup Forecast of on-site photovoltaic generation (kW)\nomeqref {0}\nompageref{1}
  \item [{\(\bm{\pi_{aFRR}^{+, *, \omega}}, \bm{\pi_{aFRR}^{-, *, \omega}}\)}]\begingroup Scenario-dependent reward vectors for downward and upward aFRR activation exceeding retail electricity tariff (CHF/kWh)\nomeqref {0}\nompageref{1}
  \item [{\(\bm{\pi_{aFRR}^{+, *}}, \bm{\pi_{aFRR}^{-, *}}\)}]\begingroup Reward vectors for downward and upward aFRR activation exceeding retail electricity tariff (CHF/kWh)\nomeqref {0}\nompageref{1}
  \item [{\(\bm{\pi_{aFRR}^{+}}, \bm{\pi_{aFRR}^{-}}\)}]\begingroup Reward vectors for downward and upward aFRR activation (CHF/kWh)\nomeqref {0}\nompageref{1}
  \item [{\(\bm{\pi_{Import}}, \bm{\pi_{Export}}\)}]\begingroup Retail and feed-in electricity tariff vectors (CHF/kWh)\nomeqref {0}\nompageref{1}
  \item [{\(\bm{C}\)}]\begingroup Cumulative-sum operator in the form of triangular matrix\nomeqref {0}\nompageref{1}
  \item [{\(\Delta\)}]\begingroup 15 minutes time interval (h)\nomeqref {0}\nompageref{1}
  \item [{\(\delta\)}]\begingroup 30 seconds time interval of the MPC model (h)\nomeqref {0}\nompageref{1}
  \item [{\(\eta\)}]\begingroup Charging/discharging efficiency of the battery (\%)\nomeqref {0}\nompageref{1}
  \item [{\(\hat{P}\)}]\begingroup Dispatch plan reference power (kW)\nomeqref {0}\nompageref{1}
  \item [{\(\overline{B}\)}]\begingroup Maximum power rating of the battery (kW)\nomeqref {0}\nompageref{1}
  \item [{\(\pi_{Power}\)}]\begingroup Power tariff (CHF/kW)\nomeqref {0}\nompageref{1}
  \item [{\(\underline{E}, \overline{E}\)}]\begingroup Minimum and maximum battery energy limits (kWh)\nomeqref {0}\nompageref{1}
  \item [{\(E_{Nom}\)}]\begingroup Nominal energy capacity of the battery (kWh)\nomeqref {0}\nompageref{1}
  \item [{\(L, \hat{L}\)}]\begingroup Net electrical load of the building and its forecast (kW)\nomeqref {0}\nompageref{1}
  \item [{\(p_\omega\)}]\begingroup Probability of occurrence of scenario \(\omega\)\nomeqref {0}\nompageref{1}
  \item [{\(P_{Transformer}\)}]\begingroup Rated power of the transformer (kW)\nomeqref {0}\nompageref{1}
  \item [{\(SOE(0)\)}]\begingroup Initial state-of-energy of the battery (kWh)\nomeqref {0}\nompageref{1}
  \item [{\(SOE(T)^\omega\)}]\begingroup Final battery state-of-energy for scenario \(\omega\) (kWh)\nomeqref {0}\nompageref{1}
\nomgroup{V}
  \item [{\(\bm{B_L^{+}}, \bm{B_L^{-}}\)}]\begingroup Battery charging and discharging power vectors for local services (kW)\nomeqref {0}\nompageref{1}
  \item [{\(\bm{B_L}\)}]\begingroup Active battery output for local services (kW)\nomeqref {0}\nompageref{1}
  \item [{\(\bm{B_{aFRR}^{+, \omega}}, \bm{B_{aFRR}^{-, \omega}}\)}]\begingroup Scenario-dependent reserved battery charging and discharging power vectors for aFRR services (kW)\nomeqref {0}\nompageref{1}
  \item [{\(\bm{B_{aFRR}^{+}}, \bm{B_{aFRR}^{-}}\)}]\begingroup Reserved battery charging and discharging power vectors for aFRR services (kW)\nomeqref {0}\nompageref{1}
  \item [{\(\bm{c^{\omega}}\)}]\begingroup Binary variable indicating battery charging or discharging for scenario \(\omega\)\nomeqref {0}\nompageref{1}
  \item [{\(\bm{P^{+}}, \bm{P^{-}}\)}]\begingroup Residual electricity consumption and surplus electricity generation vectors (kW)\nomeqref {0}\nompageref{1}
  \item [{\(\bm{SOE^{\omega}}\)}]\begingroup Battery state-of-energy for scenario \(\omega\) (kWh)\nomeqref {0}\nompageref{1}
  \item [{\(\mathbb{E}[R_{Regulation}^{'}]\)}]\begingroup Expected revenue from aFRR services (CHF)\nomeqref {0}\nompageref{1}
  \item [{\(B\)}]\begingroup Battery active power (kW)\nomeqref {0}\nompageref{1}
  \item [{\(B_0\)}]\begingroup First control action of the MPC solution (kW)\nomeqref {0}\nompageref{1}
  \item [{\(C_{Electricity}, C_{Energy}, C_{Power}\)}]\begingroup Total electricity cost and its energy and power charge components (CHF)\nomeqref {0}\nompageref{1}
  \item [{\(e\)}]\begingroup Dispatch plan tracking error (kWh)\nomeqref {0}\nompageref{1}
  \item [{\(P\)}]\begingroup Total electrical power at the utility meter (kW)\nomeqref {0}\nompageref{1}
  \item [{\(P_{Peak-shave}\)}]\begingroup Maximum power at the grid connection point (kW)\nomeqref {0}\nompageref{1}
  \item [{\(R_{Regulation}\)}]\begingroup Daily revenues from aFRR services (CHF)\nomeqref {0}\nompageref{1}
\end{thenomenclature}

\section{Introduction}\label{sec1}
\subsection{Background and Motivations}
%% Labels are used to cross-reference an item using \ref command.
The integration of renewable energy sources into power grids requires flexibility across all levels, from demand-side management to power generation and grid assets. Battery energy storage systems (BESSs) are a promising technology to support this integration, and their adoption has increased in recent years due to declining costs \citep{namor2018control, hanif2022multi}. In behind-the-meter settings with local renewable generation, such as commercial and residential buildings equipped with photovoltaic (PV) systems, the phase-out of preferential feed-in tariffs for renewable energy has further motivated the adoption of BESSs to increase self-consumption. However, the initial investment required for a single application remains relatively high. Numerous studies indicate that the profitability of BESSs depends on their ability to provide multiple services \citep{shi2018, engels2020, almasalma2020simultaneous}.

This economic potential aligns with the growing needs of grid operators for flexible resources to successfully integrate larger shares of renewable generation, both at the transmission-system level in terms of balancing power and at the distribution-grid level for congestion management and voltage control (e.g., \cite{hirth2015balancing,ali2025large}). These needs are reflected in higher market prices, most notably for balancing power, while distribution-grid services are typically not yet remunerated due to the lack of local flexibility markets, although they may be in the future. Provided that behind-the-meter assets can access balancing markets, this represents an opportunity to increase their profitability while simultaneously supporting grid operation. In this context, behind-the-meter flexible assets may even exhibit superior economic performance compared to dedicated front-of-the-meter assets, as they can inherently provide multiple services, both locally and to the grid. Consequently, the need for robust strategies capable of handling both behind-the-meter and front-of-the-meter services is evident and a relevant research question. 

This paper presents a practical and rigorous methodological toolchain, supported by extensive experimental validation, to scheduling and operation of a BESS to simultaneously provide behind-the-meter services and balancing power to the power grid cost-optimally. The proposed approach combines a two-stage convex optimization framework, for day-ahead scheduling and real-time control, with a forecasting pipeline for behind-the-meter power flows.

\subsection{Related work}\label{sec:Related_work}
Numerous strategies for front- and behind-the-meter BESS providing multiple services have been proposed in the literature. Maximizing the profitability of BESS operation requires coordinated scheduling and real-time control strategies that explicitly account for uncertainties in net-demand realizations and market conditions. This challenge is commonly addressed through a multi-layer architecture that combines (i) a scheduling layer, which optimizes the state of charge (SOC) ahead of operation, and (ii) a real-time control layer, which ensures the delivery of contracted services during operation.
Several works adopt such multi-layer or multi-timescale frameworks. In \citep{hanif2022multi}, robust programming is used to manage market uncertainties within a multi-service procurement framework for a utility-scale BESS. The scheduling stage is complemented by a sensitivity-based real-time control algorithm to accommodate each service. Similarly, \citep{namor2018control} proposes a two-phase control framework consisting of a period-ahead planning stage that allocates energy capacity to each service and a real-time stage in which control setpoints are calculated and superimposed. Extending this concept, \citep{gerini2022} introduces a three-layer structure—day-ahead, intraday, and real-time—for grid-forming inverter-interfaced BESSs, enabling simultaneous feeder dispatch and provision of frequency containment reserve and voltage control.
Multi-service operation across different time scales has also been addressed through problem decomposition techniques. In \citep{su2022}, a three-level model is proposed for a BESS simultaneously performing energy arbitrage, peak shaving, and frequency regulation. The original problem is decomposed into three subproblems, demonstrating electricity bill savings for both prosumers and consumers. In a distribution grid context, \citep{almasalma2020simultaneous} develops a control framework for residential PV-battery systems to support distribution grid voltage and provide frequency containment reserves, validated through extensive scenario-based simulations.
The economic advantages of coordinated multi-service operation have been emphasized in several studies. For example, \citep{shi2018} shows that joint optimization of peak shaving and frequency regulation yields higher savings than independent applications, reducing electricity bills for large customers by up to 12\%. Stochastic optimization approaches have also been explored: \citep{engels2020} proposes a method to optimally combine peak shaving and frequency control, which is extended in \citep{claessens2019} through the inclusion of a linear recharging policy to mitigate SOC constraint violations.
Distinct from conventional multi-stage approaches, \citep{chen2025} formulates revenue stacking from a behind-the-meter BESS as a time-average stochastic optimization problem and proposes a two-timescale online scheme based on Lyapunov optimization.
Beyond optimization-based approaches, recent research increasingly incorporates machine learning techniques. In \citep{mueller2025comparison}, imitation learning and reinforcement learning methods are shown—via simulation—to outperform rule-based scheduling and achieve performance comparable to model predictive control for residential PV-coupled BESS. The work in \citep{cardo2025deep} proposes a deep reinforcement learning strategy for renewable generation coupled with BESS participation in multiple electricity markets, demonstrating improved economic performance compared to conventional approaches. Furthermore, \citep{selim2025day} introduces a growth-optimizer-based day-ahead scheduling framework within a cyber-physical–social system, reporting improvements in cost efficiency, reliability, and battery lifetime under renewable uncertainty.

\subsection{Contribution}
Despite the breadth of optimization and control strategies proposed in the literature, comprehensive experimental validation of such multi-service frameworks remains relatively scarce. In particular, real-world implementation and validation are essential to assess performance under realistic operational constraints and uncertainties.
In contrast, this paper presents a practical and rigorous methodological toolchain supported by extensive experimental validation. The proposed framework consists of a two-stage convex optimization architecture (embedding a scheduling layer for multi-service allocation, and real-time control for service tracking and constraint enforcement); and a forecasting pipeline for the power flows of the involved behind-the-meter resources.
The real-time control formulation builds upon the approach in \citep{sossan2016} and is extended here to enable the provision of automatic frequency restoration reserve (aFRR), which constitutes one of the original contributions of this work.
The methodology is applied to the simultaneous provision of behind-the-meter services—namely peak shaving and photovoltaic self-consumption—together with aFRR participation in the grid.

The remainder of this paper is organized as follows. Section \ref{sec:Problem_Statement} presents the use case and services considered. Section \ref{sec:Problem_Formulation} details the optimal BESS scheduling problem. Section \ref{sec:Implementation_setup} presents the experimental setup. Section \ref{sec:Results_and_Discussion} discusses the experimental results. Finally, Section \ref{sec:Conclusion} concludes the paper.

\section{Problem Statement}\label{sec:Problem_Statement}
\subsection{Use case and notation}\label{subsec1}

\begin{figure}[t!]
\centering%% For centre alignment of image.
\includegraphics[width=1\linewidth]{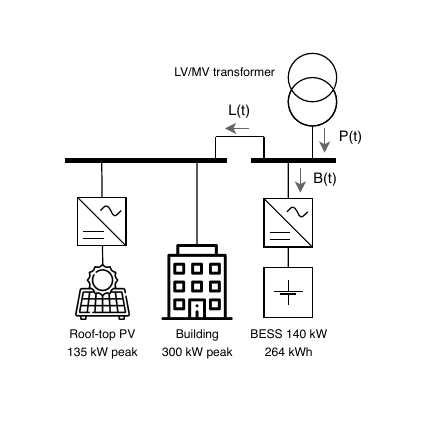}
\caption{Schematic representation of the system setup showing the physical connections and power flow.}\label{fig.setup}
\end{figure}

We consider a residential or commercial building equipped with a local BESS, roof-top PV generation, and a connection to the local electrical utility, as depicted in Fig.~\ref{fig.setup}. The grid connection point to the utility is defined at the substation transformer, where the utility meter used for electricity billing is assumed to be installed.

The building net load, defined as the electrical demand minus the photovoltaic generation, is denoted by $L(t)$, where $t$ indexes a time interval of duration $\Delta$. The time series $L(t)$, for $t = 0, 1, \dots T-1$, as well as the other signals introduced in the following, are assumed to be piecewise constant over each time interval. The active power of the BESS is denoted by $B(t)$, while the total power at the utility meter is denoted by $P(t)$. Positive values correspond to consumption, whereas negative values correspond to generation. The charging and discharging power $B(t)$ of the BESS is controllable within the limits imposed by the converter rated power and the available energy. Its optimal control to minimize operating costs is the main objective of this paper. In contrast, the load and photovoltaic generation are non-controllable; their realizations, denoted by $L(t)$, are stochastic and can only be forecasted.

Although reactive power could also be controlled, it is not considered in this study. In low-voltage distribution grids, reactive power control is only marginally effective for voltage regulation, and other connected loads are assumed to operate at power factors close to unity. Under these conditions, reactive power compensation is therefore expected to have limited relevance.

Assuming negligible cable impedances and neglecting grid losses, the power exchanged at the grid connection point during each time interval is given by the sum of the net load $L(t)$ and the battery power $B(t)$ and reads as:
\begin{align}\label{eq:P_GCP}
P(t) = B(t) + L(t).
\end{align}

The problem addressed in this paper is the scheduling and control of the battery energy storage system in order to minimize operational costs. In this setting, cost minimization is achieved through two complementary mechanisms: reducing the electricity bill and maximizing revenues by providing ancillary services to the grid. These services are defined in the following sections.

Throughout the paper, and for the sake of compact notation, unless otherwise stated, boldface symbols $\bm{x} = [x_0, x_1, \dots, x_{T-1}]$ are used to denote sequences of piecewise constant scalar values over time. For example, $\bm{B}$ represents the charging and discharging power trajectory of the battery energy storage system.

\subsection{Electricity bill reduction}
The electricity cost $C_{\mathrm{Electricity}}$ for large consumers is generally decomposed into two principal components, namely the energy charge $C_{\mathrm{Energy}}$ and the power charge $C_{\mathrm{Power}}$. The energy charge corresponds to the total electrical energy consumed over the billing period and is computed using a unit energy tariff expressed in monetary units (e.g., CHF) per kilowatt-hour. In contrast, the power charge is determined by the maximum power demand recorded during the same period. This demand is generally defined as the highest average power consumption over a 15-minute interval and is billed using a tariff expressed in CHF per kilowatt. Accordingly, the total electricity cost can be expressed as
\begin{equation}\label{eq:electricity_bill}
    C_{Electricity} = C_{Energy} + C_{Power}.
\end{equation}

The energy cost component is computed using asymmetric tariffs ${\pi}_{\mathrm{Import}}$ and ${\pi}_{\mathrm{Export}}$, applied to the residual electricity consumption $P^{+}$ and the surplus generation $P^{-}$ exported to the grid, respectively. These prices are assumed to be known in advance, as is typically the case for retail electricity tariffs, or to be forecasted, for example for consumers exposed to wholesale electricity markets.

Assuming a one day long scheduling horizon and possibly time varying electricity tariffs encoded in the corresponding price vectors, the energy cost over one day is expressed as follows:
\begin{equation}\label{eq:energy_charge}
    C_{Energy} = \Delta \cdot \left( \bm{\pi_{Import}}^\top \cdot \bm{P^+} - \bm{\pi_{Export}}^\top \cdot \bm{P^-}\right),
\end{equation}
where $\Delta$ is duration of the time interval in hours. The power charge over the scheduling horizon is defined as:
\begin{equation}\label{eq:peak_demande_charge}
    C_{Power} = \max \left( \bm{P^+} \right) \cdot \pi_{Power},
\end{equation}
where $\pi_{\mathrm{Power}}$ denotes the power tariff.

\subsection{Automatic Frequency Restoration Reserve (aFRR)}
BESSs can support aFRR, or secondary regulation, by responding to up- and down-regulation signals in real-time, with update intervals of few seconds. Usually, frequency regulation involves day-ahead or hourly capacity market and a real-time market. Resources providing this service are required to reserve some of their capacity to be made available to the system operator who can use it whenever needed; then, the balancing service provider receives a per-MW remuneration from the system operator for keeping its resources in stand-by mode during the contracted period. If a mismatch exists between the instructed dispatch and the actual resource response during the frequency regulation procurement period, penalties are applied to the balancing service provider \citep{shi2018}. Accessing this regulation market currently requires a minimum capacity, such as 1 MW in several European countries and 5 MW in Switzerland \citep{entsoe2019}.

As behind-the-meter BESS typically have smaller capacities than the minimum entry requirements of frequency regulation markets, we assume that rewards may be received when the BESS is activated to provide frequency regulation, even if it does not participate in the capacity market. We further assume that the system behaves as a price-taker and that, for the purposes of this study, no penalties related to the non-provision of services are considered. Behind-the-meter resources may also be aggregated by aggregators; in such cases, different market rules may apply and penalties for non-provision of services may be enforced.

Although it is assumed that BESS capacity is not explicitly committed in the aFRR market, the optimal allocation of BESS energy capacity and power remains beneficial to ensure that sufficient state-of-charge reserves are available in real time to capitalize on favorable balancing market prices. Since the amount of battery capacity to be reserved for this service in future hours depends on the regulating price, which is not known during the planning phase, price forecasts are therefore used to support optimal scheduling.

The daily revenue $R_{Regulation}$ from providing balancing power is modeled as:
\begin{equation}\label{eq:daily_aFRR_reward}
    R_{Regulation} = \Delta \cdot \left( {\bm{\pi_{aFRR}^{+}}}^\top \cdot \bm{B_{aFRR}^+} + {\bm{\pi_{aFRR}^{-}}}^\top \cdot \bm{B_{aFRR}^{-}} \right)
\end{equation}
where $\bm{B_{aFRR}^+}$ and $\bm{B_{aFRR}^-}$, both non negative, denote the reserved charging and discharging powers of the battery for the provision of down and up regulation, respectively, and $\bm{\pi_{aFRR}^+}$ and $\bm{\pi_{aFRR}^-}$ denote the corresponding rewards from the system operator. The reserved charging and discharging powers are computed using the methodology described in the following section, whereas the rewards are provided as inputs to the optimization problem based on price forecasts.

\section{Problem Formulation}\label{sec:Problem_Formulation}
We consider the problem of determining and actuating a BESS schedule that minimizes the operational costs of a smart building equipped with a behind-the-meter BESS and photovoltaic generation. To this end, a two-stage methodology is proposed. In the first stage, day-ahead scheduling is used to allocate BESS energy capacity and power levels to the most profitable services, including local services and the provision of balancing power. In the second stage, real-time control adjusts the BESS power output to track the scheduled operation and to respond to real-time events, such as forecast uncertainties and their realization. These two stages are described in the following sections.

\subsection{Day-ahead scheduling}
This stage is computed on the day before operation and aims to schedule BESS utilization so as to minimize operational costs or, equivalently, maximize profit. Specifically, two BESS charging and discharging time series are determined: one associated with the provision of local services and one associated with the provision of balancing services. These time series are selected to minimize costs or maximize revenues arising from local service provision and regulating power. Both time series must satisfy the power and energy capacity constraints of the BESS, ensuring that their actuation does not violate operational limits. The resulting problem is formulated as a constrained techno-economic optimization that minimizes the total cost while respecting BESS operational constraints, as detailed below. Before introducing the optimization problem, the models adopted for the cost of operation and the BESS dynamics are presented.

\subsubsection{Energy cost model}
From \eqref{eq:P_GCP}, the power at the grid connection point is the sum of the buildings’ net load and the BESS output. Hence, \eqref{eq:energy_charge} can be written as:
\begin{equation}\label{eq:energy_cost_model}
\begin{split}
C_{Energy} = \Delta & \Bigl( \bm{\pi_{Import}}^\top \left[ \bm{\hat{L}} + \bm{B_{L}^+} - \bm{B_{L}^-} \right]^+ + \\
& - \bm{\pi_{Export}}^\top \left[ \bm{\hat{L}} + \bm{B_{L}^+} - \bm{B_{L}^-} \right]^- \Bigr)
\end{split}
\end{equation}
where $\left[\cdot\right]^+$ and $\left[\cdot\right]^-$ are the positive and negative part operators, respectively, $\bm{\hat{L}}$ is the day-ahead net-load forecast, and $\bm{B_{L}^+}$ and $\bm{B_{L}^-}$ are, respectively, the non-negative charging and discharging power capacities reserved for local services. These are derived from historical disaggregated measurements of the building load and PV forecast, following the approach described in section \ref{sec:DA-net:load_forecast}.

Eq.~\eqref{eq:energy_cost_model} is convex under the assumption that \( \left( \bm{\pi}_{\text{Import}} - \bm{\pi}_{\text{Export}} \right) \) is non-negative, as proven in \ref{app:convex_reformulation_cost_function}. This condition is generally satisfied, since the retail electricity tariff is typically higher than the feed-in tariff due to grid charges.

% \begin{equation}\label{eq:energy_cost_model_reformulated}
% \begin{split}
% C_{Energy} = \Delta & \Bigl( \left( \bm{\pi_{Import}} - \bm{\pi_{Export}} \right)^\top \bigl[ \bm{\hat{L}} + \bm{B_{L}^+} - \bm{B_{L}^-} \bigr]^+ + \\
% & + \bm{\pi_{Export}}^\top \bigl( \bm{\hat{L}} + \bm{B_{L}^+} - \bm{B_{L}^-} \bigr) \Bigr)
% \end{split}
% \end{equation}

It is also worth remarking that, since the electricity tariff for consumption is typically higher than the feed-in tariff for power injected into the grid, the proposed formulation naturally promotes photovoltaic self-consumption.

\subsubsection{Revenues from balancing services}
\begin{figure}[t!]%% placement specifier
\centering%% For centre alignment of image.
\includegraphics[width=1\linewidth]{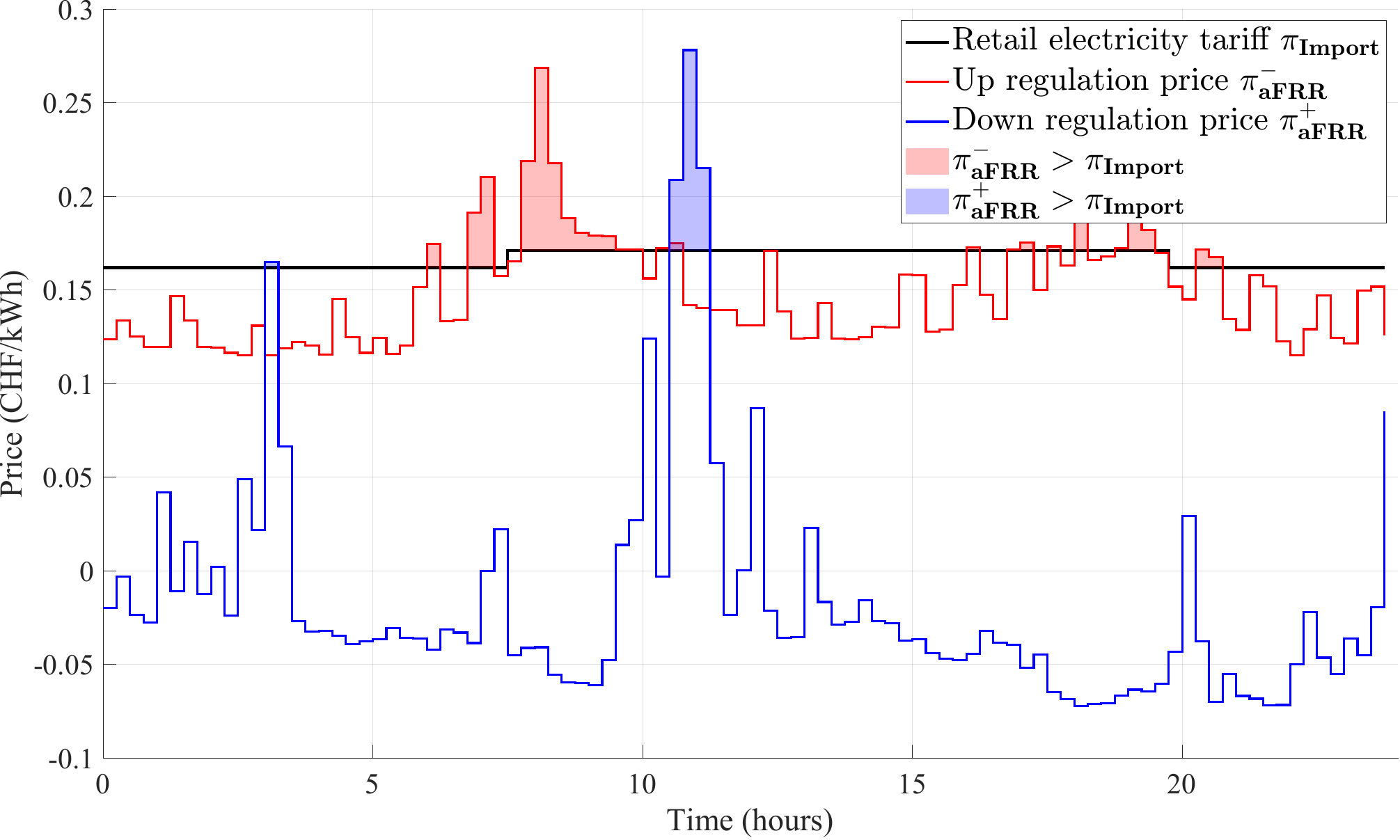}
\caption{Electricity tariffs and prices for upward and downward regulations throughout the day.}\label{fig.pi_afrr_star}
\end{figure}

Prices for upward and downward regulation may be positive or negative, indicating that payments may flow from the system operator to the balancing service provider when the latter is remunerated for its services, or conversely from the balancing service provider to the system operator when the former is willing to pay to avoid providing the contracted power, if this is economically advantageous based on look-ahead considerations \citep{brijs2015}. An example of upward and downward regulation prices is shown in Fig.~\ref{fig.pi_afrr_star}, along with their comparison against the retail electricity tariff.

To prevent that, in the day-ahead problem, the BESS might charge by leveraging regulating power prices that are smaller than the retail electricity tariff (which might result in contractual violation with the electricity retailer), we require the BESS to respond to regulating power prices that are larger than the retail electricity tariff. To this end, we model the prices for down regulation as follows:
\begin{equation}\label{eq:prices_down_regulation}
    \bm{\pi_{aFRR}^{+, *}} =
    \begin{cases}
      \bm{\pi_{aFRR}^{+} - \pi_{Import}} & \text{if $\bm{\pi_{aFRR}^{+}} \: > \: \bm{\pi_{Import}}$}\\
      -\infty & \text{Otherwise}\\
    \end{cases}       
\end{equation}
Similarly, the prices for upward regulation are the following:
\begin{equation}\label{eq:prices_up_regulation}
    \bm{\pi_{aFRR}^{-, *}} =
    \begin{cases}
      \bm{\pi_{aFRR}^{-} - \pi_{Import}} & \text{if $\bm{\pi_{aFRR}^{-}} \: > \: \bm{\pi_{Import}}$}\\
      -\infty & \text{Otherwise}\\
    \end{cases}       
\end{equation}
Based on this, the profit from providing regulation services is modeled as:
\begin{equation}\label{eq:daily_aFRR_reward_reformulated}
    R_{Regulation} = \Delta \left( {\bm{\pi_{aFRR}^{+, *}}}^\top \bm{B_{aFRR}^+} + {\bm{\pi_{aFRR}^{-, *}}}^\top \bm{B_{aFRR}^{-}} \right)
\end{equation}

\subsubsection{Modeling Uncertainty in Balancing Power Prices}
Uncertainties in the scheduling problem arise from forecasts of the aggregated net load $\bm{\hat{L}}$ and of the prices for upward $\bm{\pi_{aFRR}^{-,*}}$ and downward $\bm{\pi_{aFRR}^{+,*}}$ regulation. The main source of uncertainty is associated with balancing power prices. For the net load $\bm{\hat{L}}$, point forecasts are adopted and a deterministic setting is therefore considered. To address uncertainty in regulating power prices, a two-stage stochastic optimization approach is employed: in the first stage, the BESS energy and power capacities are allocated to local services; in the second stage, the capacity is allocated using a scenario-based approach based on historical aFRR price data, under the assumption that these historical prices are representative of future outcomes.

In these settings, the profit from balancing services can be reformulated in terms of expected value $\mathbb{E}[R_{Regulation}^{'}]$ as:
\begin{equation}\label{eq:daily_aFRR_reward_scenarios}
\begin{split}
\mathbb{E}[R_{Regulation}^{'}] = \Delta \sum_{\omega = 1}^{\Omega} p_{\omega} & \Bigl( 
 {\bm{\pi_{aFRR}^{+*, \omega}}}^\top \bm{B_{aFRR}^{+, \omega}} +\\
& + {\bm{\pi_{aFRR}^{-*, \omega}}}^\top \bm{B_{aFRR}^{-, \omega}} \Bigr)
\end{split}
\end{equation}
where $\omega$ denotes a possible realization (scenario) among a finite number of scenarios $\Omega$, and $p_{\omega}$ refers to the probability of occurrence of scenario $\omega$. The decision variables of the second stage are sequences of power allocations for balancing services $\bm{B_{aFRR}^{+,\omega}}$ and $\bm{B_{aFRR}^{-,\omega}}$ for each scenario, depending on the aFRR prices for upward and downward reserve, $\bm{\pi_{aFRR}^{+,*,\omega}}$ and $\bm{\pi_{aFRR}^{-,*,\omega}}$.

\subsubsection{Battery model}
The state-of-charge evolution of the BESS is captured in terms of the dynamics of its state-of-energy ($SOE$) in kWh, which is modeled as:
\begin{equation}\label{eq:soe}
\begin{split}
\bm{SOE^\omega} = \bm{SOE(0)} + \Delta \bm{C} & \Biggl(
\eta \left( \bm{B_L^+} + \bm{B_{aFRR}^{+, \omega}} \right) + \\
& - \frac{1}{\eta}\left( \bm{B_L^-} + \bm{B_{aFRR}^{-, \omega}} \right) \Biggr)
\end{split}
\end{equation}
where $\bm{SOE(0)}$ is the initial state-of-charge vector of the BESS (known from measurements at the time of computing the schedule), $\eta$ is the charging/discharging efficiency (assumed symmetric), and $\bm{C}$ is the cumulative-sum in the form of a lower triangular matrix ensuring that SOE dynamics at each time t reflects all past charging/discharging actions.

The BESS operational constraints require that the battery state-of-energy ($SOE$) remains within a customizable lower and upper bound, denoted by $\underline{E}$ and $\overline{E}$, respectively, with the latter strictly smaller than the nominal BESS energy capacity $E_{\text{Nom}}$. Moreover, under the assumption that the power capability is independent of the battery $SOE$ \citep{cassano2024}, the battery power must not exceed the rated power of the converter. Formally, BESS constraints read as follows:
\begin{align}
\bm{1} \cdot \underline{E} & \le \bm{SOE^{\omega}} \le \bm{1} \cdot \overline{E} \label{eq:soe_bounds} \\
\bm{0} & \le \bm{B_L^+} + \bm{B_{aFRR}^{+, \omega}} \le \bm{c^\omega} \cdot \overline{B} \label{eq:charging_power_bounds} \\
\bm{0} & \le \bm{B_L^-} + \bm{B_{aFRR}^{-, \omega}} \le \left(\bm{1} - \bm{c^\omega} \right) \cdot \overline{B} \label{eq:discharging_power_bounds}
\end{align}
where $\overline{B}$ is the maximum power rating of the BESS and the binary decision variable $\bm{c^\omega} \in \{0,1\}$ is introduced to ensure mutual exclusive charging and discharging of the battery.

\subsubsection{Maximum power at the grid connection point and peak shaving}
The term $C_{\mathrm{Power}}$, introduced in Eq.~\eqref{eq:peak_demande_charge} to account for peak power tariff reduction, includes a max operator. This nonlinearity is handled by introducing an auxiliary decision variable $P_{\mathrm{Peak\text{-}shave}}$ (see \ref{app:linearize_max_operator}), defined as follows:
\begin{equation}\label{eq:peak_shaving_constraint}
   \bm{1} \cdot P_{Peak-shave} \ge \bm{\hat{L}} + \bm{B_L^+} - \bm{B_L^-}
\end{equation}

When multi-tasking the battery, it is assumed that a prior agreement has been established with the grid operator for the provision of balancing services. In addition to remuneration for these services, this agreement allows the exclusion of balancing power from the prosumer’s electricity bill. Accordingly, Eq.~\eqref{eq:peak_shaving_constraint} enforces that the power exchanged with the grid, excluding balancing power, remains below a peak-shaving threshold, which is determined as a decision variable of the optimization problem. Moreover, the total power exchange with the grid, including balancing power, must not exceed the transformer rated power, as expressed in the following equation:
\begin{equation}\label{eq:transformer_constraint}
    \bm{1} \cdot P_{Transformer} \ge \bm{\hat{L}} + \bm{B_L^+} - \bm{B_L^-} + \bm{B_{aFRR}^{+, \omega}} - \bm{B_{aFRR}^{-, \omega}}
\end{equation}

\subsubsection{Formulation of the scheduling problem}
Having introduced the main models for the costs and the BESS, we are now ready to formulate the optimal scheduling problem to dispatch the smart building campus.

The scheduling problem aims to optimize the operational costs of the smart building by minimizing electricity costs and maximizing profits from aFRR, subject to system and peak shaving constraints. The problem is formulated as:
\begin{subequations}\label{eq:DA-problem}
\begin{align}\label{eq:objective_DA-problem}
\begin{aligned}
\min_{\substack{\bm{B^{+}_{L}}, \bm{B^{-}_{L}},\\ \bm{B^{+, \omega}_{aFRR}}, \bm{B^{-, \omega}_{aFRR}},\\ \bm{c^\omega}, P_{\text{Peak-shave}}}}
\Bigg\{& C_{Energy} (\bm{B^{+}_{L}}, \bm{B^{-}_{L}}) + C_{Power} \left( \bm{B^{+}_{L}}, \bm{B^{-}_{L}} \right) + \\
& +\mathbb{E}\left[R_{Regulation}^{'}\left( \bm{B^{+, \omega}_{aFRR}}, \bm{B^{-,\omega}_{aFRR}} \right)\right] \Bigg\}
\end{aligned}
\end{align}
subject to:
\begin{align}
& \bm{0} \le \bm{B_L^{+}} + \bm{B_{aFRR}^{+, \omega}} 
  \le \bm{c^\omega} \cdot \overline{B} 
  \label{eq:bess_charging_DA_problem} \\
& \bm{0} \le \bm{B_L^{-}} + \bm{B_{aFRR}^{-, \omega}} 
  \le \bigl( \bm{1} - \bm{c^\omega} \bigr) \cdot \overline{B} 
  \label{eq:bess_discharging_DA_problem} \\
&\begin{aligned}
\bm{SOE^{\omega}} = \bm{SOE(0)}
& + \Delta \bm{C} 
   \eta \left( \bm{B_L^{+}} + \bm{B_{aFRR}^{+, \omega}} \right) + \\
& \quad - \Delta \bm{C} 
   \frac{1}{\eta} \left( \bm{B_L^{-}} + \bm{B_{aFRR}^{-, \omega}} \right)
\end{aligned}
\label{eq:bess_soe_DA_problem} \\
& \bm{1} \cdot \underline{E} \le \bm{SOE^{\omega}} \le \bm{1} \cdot \overline{E}
  \label{eq:bess_soe_bounds_DA_problem} \\
& \bm{1} \cdot P_{Peak-shave} \ge \bm{\hat{L}} + \bm{B_L^+} -     \bm{B_L^-} 
    \label{eq:peak_shaving_DA_problem} \\
& \bm{1} \cdot P_{Transformer} \ge \bm{\hat{L}} + \bm{B_L^+} - \bm{B_L^-} + \bm{B_{aFRR}^{+, \omega}} - \bm{B_{aFRR}^{-, \omega}}
    \label{eq:transformer_DA_problem}
\end{align}

Finally, to prevent starting the next day with a fully saturated/discharged battery, the constraint \eqref{eq:soe_final} enforces some residual energy in the battery at the end of the day.
\begin{align}\label{eq:soe_final}
    SOE(T)^{\omega} = SOE(0)
\end{align}
\end{subequations}

This day-ahead scheduling problem is a mixed integer linear program (MILP) which can be solved with off-the-shelf optimization libraries. It is executed every day at midnight for the next 24 hours, in 15-minute intervals. The output of the problem is the expected charging/discharging trajectories for the BESS. The BESS power, combined with the point predictions of the net load, constitutes a power flow time series at the grid connection, which we call the dispatch plan. Tracking this dispatch plan ensures the minimization of the predicted cost of electrical operations and peak shaving for the smart building complex. During real-time, the BESS is controlled, as discussed in the next section, so as to track the dispatch plan in an effort to attain the predicted minimum costs of operation. The problem is implemented in Matlab, and the optimization is solved using the Gurobi solver.

\subsection{Real-time operation}
At midnight on the day of operation, the dispatch plan enters into force for the subsequent 24 hours. The BESS is operated to ensure that the realized active power at the grid connection point tracks the dispatch plan as accurately as possible, on a best-effort basis, depending on the accuracy of the net load forecast. In addition, the BESS contribution is activated to provide aFRR and to maximize revenues from aFRR provision when favorable prices are available.

The control strategy is implemented using a rolling-horizon model predictive control (MPC) scheme, with a twofold objective: (i) minimizing the mismatch between the average net load and the dispatch plan over each 15-minute interval, and (ii) providing aFRR services subject to the available BESS capacity. The control problem is re-solved every 30 seconds. The formulation is inspired by the approach proposed in \citep{sossan2016} and, as an original contribution of this paper, is extended to allow for the provision of aFRR.

In the following formulation, time evolution is organized into two nested time grids. The index $k = 0, 1, 2, \ldots, K-1$ denotes the 30-second rolling time steps, where $K=30$ is the number of 30-second intervals within a 15-minute period. The index $t = 0, 1, 2, \ldots, T-1$ denotes the 15-minute intervals over a 24-hour horizon, with $T = 96$.

Real-time quantities are therefore indexed using a double index, such as the BESS active power $B(t,k)$, where $t$ denotes the 15-minute interval and $k$ the corresponding 30-second subinterval. For intuitive interpretation, this can be viewed as storing measurements in matrix form, where each row corresponds to a 15-minute interval $t$ and each column corresponds to a 30-second subinterval $k$ within $\Delta$. Averaging over the columns yields the mean value over the 15-minute interval, enabling an intuitive interpretation of the MPC real-time control action as achieving the dispatch plan at a 15-minute resolution through finer-grained control actions at a 3-second resolution.

\subsubsection{Dispatch plan tracking}
The primary role of the MPC is to steer the BESS real-power injection so as to ensure compliance with the day-ahead dispatch plan at a 15-minute resolution.

In other words, rather than enforcing instantaneous power matching, which would be impractical due to actuation and communication latencies, the MPC regulates the BESS such that any power imbalance is minimized, and ideally eliminated, by the end of each 15-minute interval. In practice, this corresponds to enforcing that the average net power measured over each 15-minute interval coincides with the prescribed dispatch target. The formulation of the tracking error is as follows.

First, the dispatch plan value for time interval $t$ is extracted from the database and denoted by $\hat{P}(t,k)$, where $\hat{P}(t,k)$ represents a piecewise-constant signal indexed by $k = 0,1,\dots, K-1$. The tracking error at any time interval $(t,k)$ is given by
\begin{align}\label{eq:energy_error0}
    e(t,k) & = \delta \left( \hat{P}(t,k) - P(t,k)\right) = \\
    & = \delta \left( \hat{P}(t,k) - L(t,k) - B_L(t,k)\right), \label{eq:energy_error1}
\end{align}
where, in \eqref{eq:energy_error1}, the power at the grid connection point is decomposed into contributions from the net load and the BESS.

Assuming now to be at time interval $(t,k^\star)$, the accumulated dispatch error is given by (neglecting the constant $\delta$, that acts on all terms):
\begin{align}\label{eq:energy_error}
\begin{aligned}
\sum_{k=0}^{K-1} e(t,k) =& \sum_{j=0}^{k^\star} \left( \hat{P}(t,j) - P(t,j) \right) + \\
&+ \sum_{j=k^\star+1}^{K-1} \left( \hat{P}(t,j) - \hat{L}(t,j) - B_L(t,j) \right).
\end{aligned}
\end{align}
where the first term represents the already occurred realization (e.g., $P(t,j)$ is a measurement), while the second term represents the future realization, with $\hat{L}(t,j)$ denoting near-real-time net-demand forecasts and $B_L(t,j)$ the battery control action to be computed by the MPC.
To achieve successful dispatch, the tracking error $\sum_{k=0}^{K-1} e(t,k)$ should be zero, or as small as possible, providing a natural interpretation as a tracking performance metric to be minimized in control or MPC problem (e.g., in a norm-2 or absolute value sense). Rearranging the former expression yields:
\begin{align}
    K \hat{P}(t,k) -
    \sum_{j=0}^{k^\star} P(t,j) - 
    \sum_{j=k^\star+1}^{K-1} \hat{L}(t,j) - 
    \sum_{j=k^\star+1}^{K-1} B_L(t,j).
\end{align}
This last expression highlights that the tracking error can be written as
\begin{align}\label{eq:tracking_error_asbtract_form}
    \sum_{k=0}^{K-1} e(t,k) = a - b^\top \bm{B_L(t,k)},
\end{align}
where the scalar $a$ collects all known terms (including the dispatch plan, measurements of the power at the grid connection point, and net-demand forecasts), and $\bm{B_L}$ denotes the vector of BESS charging and discharging control actions over the remainder of the interval, i.e., over a horizon that shrinks within each 15-minute interval. In energy terms, this cumulative BESS contribution over the remaining horizon is denoted by:
\begin{equation}
    \delta \cdot b^\top \bm{B_L(t,k)} = \delta \cdot b^\top \left( \bm{B_L^+(t,k)} - \bm{B_L^-(t,k)}\right)
\end{equation}
where $\bm{B_L}$ is decomposed to non-negative vector terms $\bm{B_L^+}$ and $\bm{B_L^-}$ representing charging and discharging actions, respectively.

Equation~\eqref{eq:tracking_error_asbtract_form} is linear in the decision variable $\bm{B_L}$. Minimizing its absolute value therefore yields a convex objective function, which can be efficiently integrated into the MPC tracking problem, as discussed in the next subsection.

\subsubsection{aFRR provision}
In addition to tracking the day-ahead dispatch plan, the MPC can activate the BESS to provide aFRR services, with the objective of maximizing the associated revenues, as formulated in Eq.~\eqref{eq:daily_aFRR_reward}. In this formulation, the MPC is designed to enforce mutual exclusivity between charging and discharging actions, ensuring that the BESS responds in the same direction when pursuing both objectives.

\subsubsection{MPC formulation}
The MPC aims at minimizing the absolute deviation between the energy delivery by the battery and the energy error $e(t,k)$, while maximizing profit from the provision of aFRR and subject to system constraints. The cost function is calculated on the (shrinking) horizon $k,\dots,K-1$. By denoting with the bold typeface the decision variables over the time horizon (i.e., $\mathbf{x} = [x(t,k),x(t,k+1),\dots,x(t,K-1)$]), the cost function reads as follows:

\begin{equation}\label{eq:objective_MPC}
\begin{aligned}
\min_{\substack{\bm{B_L^+}, \bm{B_L^-},\\ 
                \bm{B_{aFRR}^{+}}, \bm{B_{aFRR}^{-}},\\ 
                \bm{c}}} \quad
& \delta \left| a - b^\top \!\left( \bm{B_L^+(t,k)} - \bm{B_L^-(t,k)} \right) \right| + \\
& \quad - R_{\text{Regulation}}\!\left(\bm{B_{aFRR}^{+}(t,k)}, \bm{B_{aFRR}^{-}(t,k)}\right).
\end{aligned}
\end{equation}

subject to:
\begin{align}
\text{Eqs.~\eqref{eq:soe} –~\eqref{eq:discharging_power_bounds}}, && \forall \text{~time intervals~} k, k+1, \dots, K.
\end{align}

The cost function $R_{Regulation}$ is as in Eq.~\eqref{eq:daily_aFRR_reward_reformulated}: positive and negative aFRR prices for the next 30 seconds are now known from the system operator and are assumed constant until $K$.

Since the objective function is linear (see \ref{app:abs_term_approximation}) and all the inequality constraints are convex, the formulation of the real-time MPC is a convex optimization problem. The MPC provides the battery trajectory for the whole residual horizon from the current $k$ until the end of the 15 minutes ($k = K-1$), however, only the first component of the solution, denoted by
\begin{align}
    B_0(t,k)=(B_L^+(t,k) - B_L^-(t,k)) + (B_{aFRR}^+(t,k) - B_{aFRR}^-(t,k))
\end{align}
is considered for actuation, and the rest of the vector is disregarded. 

\begin{figure*}[t]%% placement specifier
\centering%% For centre alignment of image.
\includegraphics[width=0.9\linewidth]{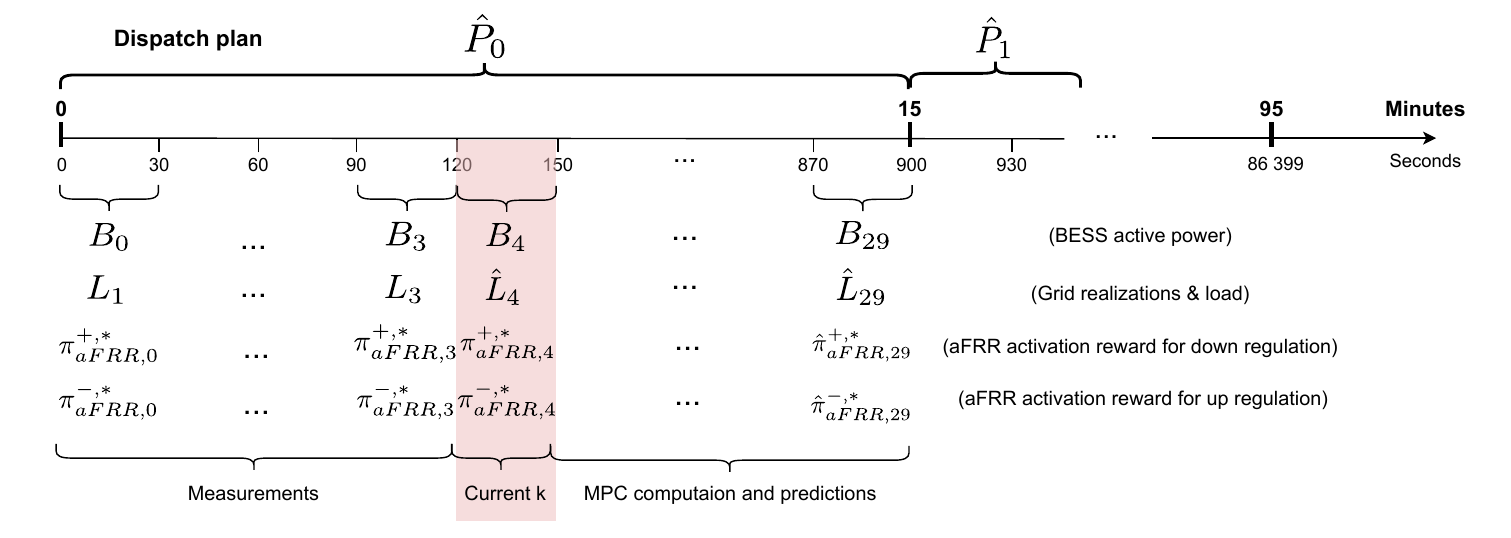}
\caption{Timeline of the real-time control illustrating the unfolding of real-time measurements and input signals with respect to the day-ahead dispatch plan.}\label{fig.timeline_MPC}
\end{figure*}

The quantities involved in the computation of the MPC control action are shown in Fig.~\ref{fig.timeline_MPC}.

\subsection{Forecasting load and balancing power prices}
Forecasting the stochastic input quantities appearing in the scheduling and real-time optimization problems is essential for optimizing the operation of the BESS and ensuring that sufficient power and energy are available when needed to maximize economic performance.

Forecasting power time series, such as electrical loads and PV generation, at low aggregation levels is becoming increasingly important due to the growing penetration of decentralized energy resources. This has immediate practical applications in contexts such as energy communities and the real-time control of distribution grids for congestion management and voltage regulation. At low aggregation levels, load diversity is typically limited; as a result, stochastic events, such as the inrush current of electric motors or sudden local losses of generation due to cloud-induced solar occlusions, can be dominant compared to the series magnitude, making the forecasting task more challenging than at higher aggregation levels (e.g., \cite{sossan2019solar, shaqour2022electrical}).

A wide variety of forecasting models have been explored in the literature, ranging from first-principles models to time-series and machine-learning approaches, and more recently extending to the forecasting of distributed load responses to flexibility-activation control signals (e.g., \cite{nespoli2025global, ding2025optimal}). The choice of the forecasting output, such as point predictions, prediction intervals, or scenarios, typically depends on the requirements of the scheduling problem, in particular on whether it is formulated in a deterministic setting or explicitly accounts for uncertainty, for instance through robust or stochastic optimization. Of particular interest in the context of control of distributed energy resources is the concept of hierarchical forecasting, which provides a reconciliation layer to ensure consistency among forecasts at different aggregation levels, for example at the grid connection point as well as at the level of individual resources (e.g., \cite{nespoli2020hierarchical}).

\subsubsection{Day-ahead net-load forecasts}\label{sec:DA-net:load_forecast}
The net-load forecast $\bm{\hat{L}}$ for the day-ahead stage is obtained by \emph{separately forecasting} the building’s gross electrical demand $\bm{\hat{L}_{\mathrm{Gross}}}$ and the on-site photovoltaic (PV) generation $\bm{\hat{P}_{\mathrm{PV}}}$, and then \emph{combining} these forecasts. The net load is defined as the difference between gross demand and PV generation,
\begin{equation}\label{eq:net-load definition}
    \bm{\hat{L}} = \bm{\hat{L}_{\mathrm{Gross}}} - \bm{\hat{P}_{\mathrm{PV}}}.
\end{equation}

Gross load forecasting is performed using a similarity-based approach applied to historical measurements. In particular, historical load time series corresponding to the same day type (working or non-working day) as the target day are first identified from a measurement database. From this subset, time series closest in time to the target day are selected and further filtered based on meteorological similarity, specifically irradiance and ambient temperature, to account for similarities in electrical space-heating demand. Once $N$ time series are selected, where $N$ is a tunable parameter, point predictions of the expected gross load are obtained by averaging the selected historical load time series $\bm{\hat{L}_{\mathrm{Gross}}^{i}}$ as follows:
\begin{equation}\label{eq:gross_demand_mean}
    \bm{\hat{L}_{\mathrm{Gross}}}
    = \frac{1}{N} \sum_{i=1}^{N} \bm{\hat{L}_{\mathrm{Gross}}^{i}} .
\end{equation}
Point predictions of PV generation are obtained by coupling numerical weather predictions with a physical model of the PV plant, as described in \cite{sossan2019solar}.

\begin{figure}[t!]%% placement specifier
\centering%% For centre alignment of image.
\includegraphics[width=1\linewidth]{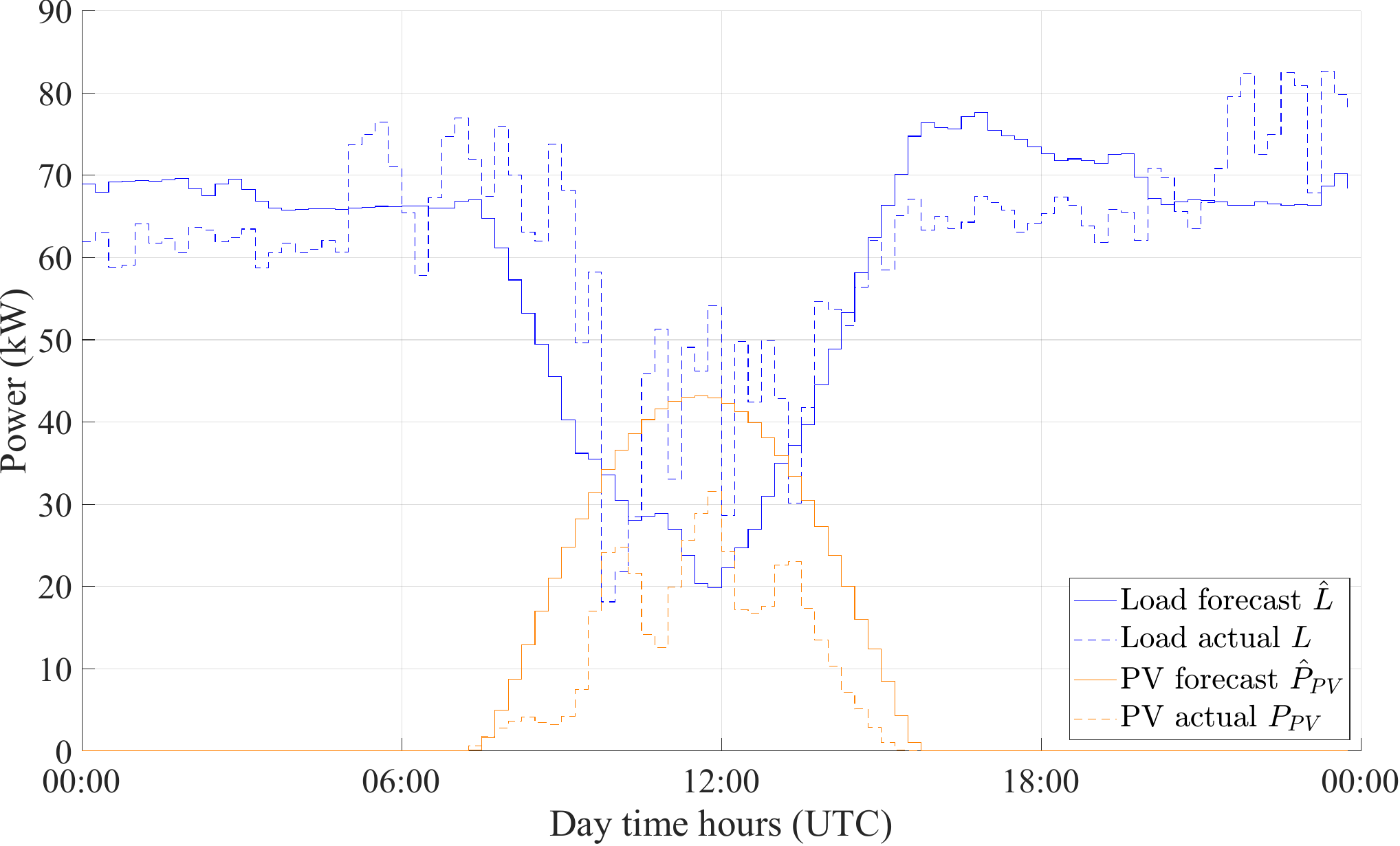}
\caption{Building net-load and PV generation time series: point predictions and realized values.}\label{fig.forecast_performance}
\end{figure}

The forecast performance of the proposed methodology is illustrated in Fig.~\ref{fig.forecast_performance}, which compares the measured and forecasted net load of the building, as well as the PV generation, over an entire day. As shown, the similarity-based gross load forecast combined with the clear-sky PV model accurately captures the overall net-load dynamics, while minor discrepancies reflect residual forecasting errors.

\subsubsection{Intra-day net-load forecasts for real-time control}\label{sec:intraday_net_load_forecast}
During real-time operation, a persistence-based approach is employed for short-term net-load forecasting, leveraging recent measurements to generate short-term predictions. This choice is motivated by the fact that, for very short forecasting horizons, persistence-based models are known to outperform more complex alternatives in practice (see, e.g., \cite{miyasawa2021spatial}). More specifically, the forecast at each MPC time step depends on its position within the current 15-minute interval:
\begin{itemize}
    \item if the current time step corresponds to the start of the 15-minute interval, the net-load forecast is set equal to the average net load over the preceding 15 minutes;
    \item if the current time step occurs after the start of the interval, the forecast is computed as the average net load from the beginning of the current 15-minute period up to the current time step.
\end{itemize}

\subsubsection{aFRR prices forecasts}
At this stage, a simple persistence model is adopted for aFRR activation prices, whereby the future aFRR activation price at each control step is assumed to be equal to its current value, which is known from the energy market. Although this model is extremely simple, it offers straightforward interpretability and robustness.

Forecasting electricity and ancillary service market prices is a complex task due to the large number of influencing factors, ranging from weather conditions to network congestion and dispatch decisions. While more advanced forecasting methods such as black-box models trained on large and heterogeneous datasets (e.g., \cite{pavirani2025predicting}) exist, their use is beyond the scope of the present study. The adoption of more sophisticated price forecasting models will be considered in future work.

\section{Implementation setup}\label{sec:Implementation_setup}
\subsection{Experimental setup}
\begin{figure}[!t]
    \centering
    \begin{subfigure}{\columnwidth}
        \includegraphics[width=\columnwidth]{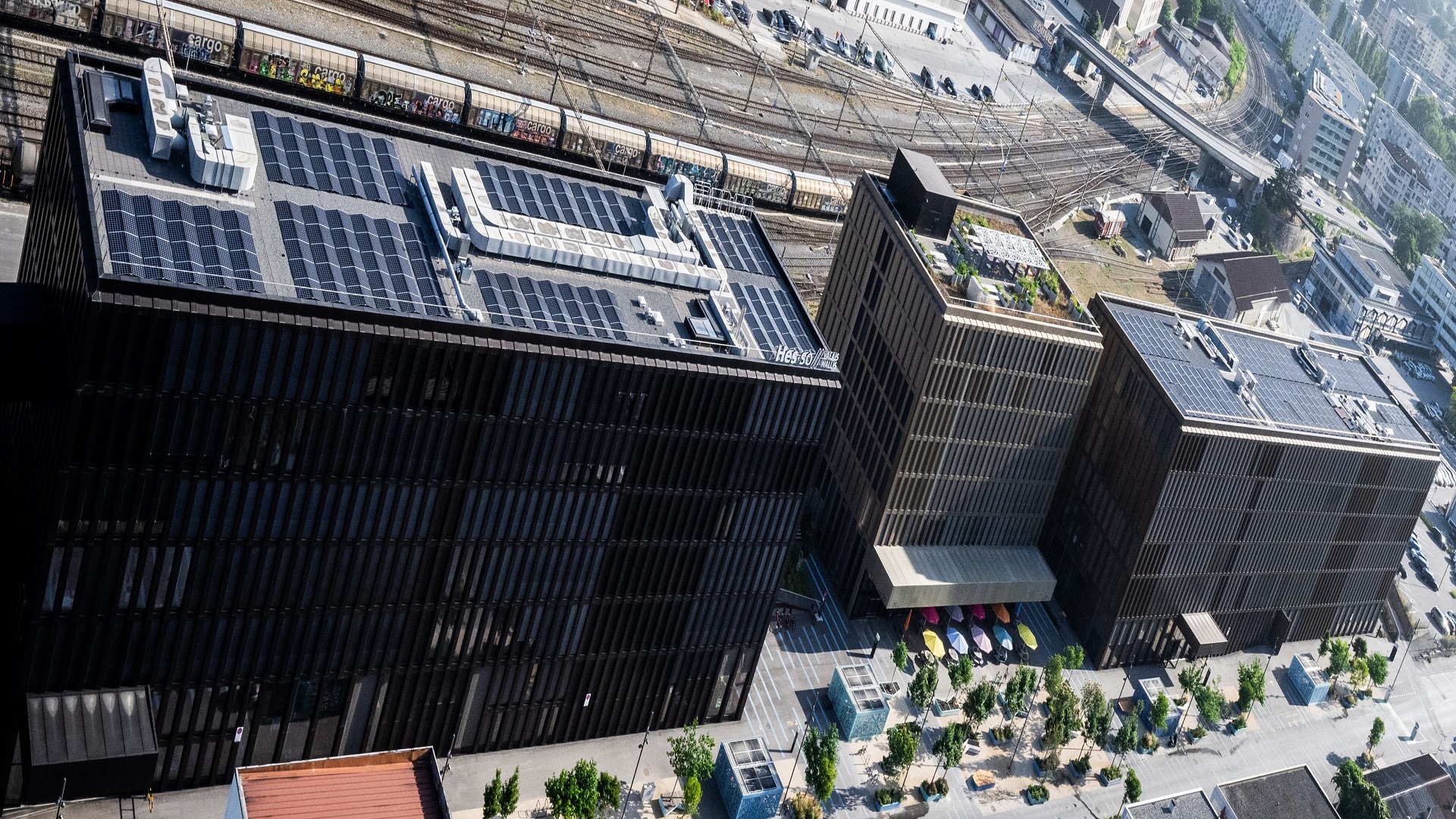}
        \caption{}
        \label{fig.EnergypolisCampus}
    \end{subfigure}

    \begin{subfigure}{\columnwidth}
        \includegraphics[width=\columnwidth]{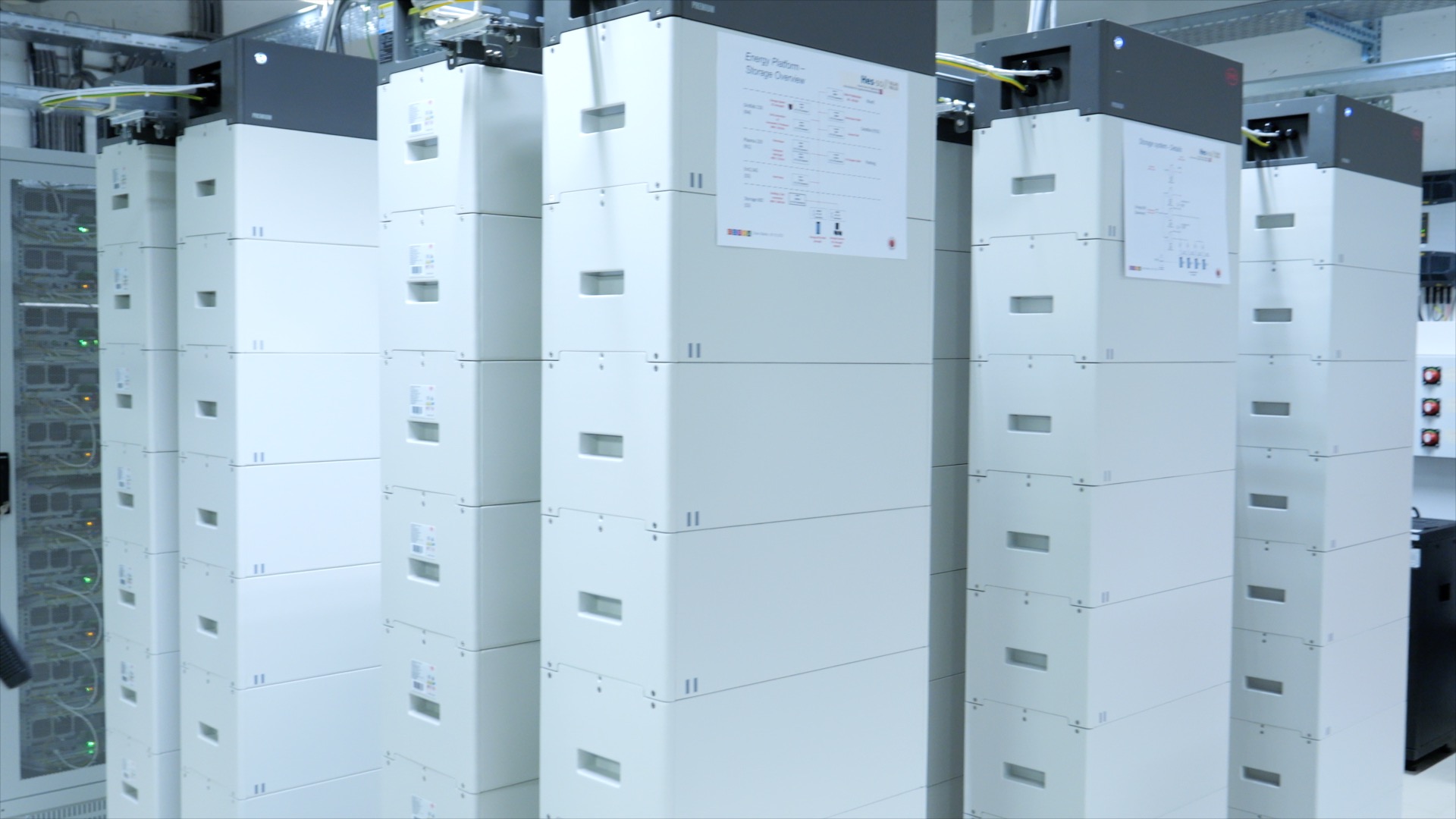}
        \caption{}
        \label{fig.BESS}
    \end{subfigure}

    \caption{Experimental setup used to validate the developed methodology: (a) the Energypolis campus of HES-SO Valais-Wallis in Sion, and (b) the BESS.}
    \label{fig.Experimental_setup}
\end{figure}

The proposed control strategy is validated in a real-life experimental setting using the infrastructure available at the Energypolis Campus of HES-SO Valais-Wallis in Sion, Switzerland. The experimental setup consists of an office building located at Rue de l’Industrie 23, 1950 Sion, equipped with a 135~kW rooftop photovoltaic (PV) installation and a 264~kWh / 140~kVA battery energy storage system (BESS) (Fig.~\ref{fig.Experimental_setup}). Tariffs and costs reported below are expressed in the local currency, Swiss francs (CHF).

\begin{figure}[!t]
    \centering
    \begin{subfigure}{\columnwidth}
        \includegraphics[width=\columnwidth]{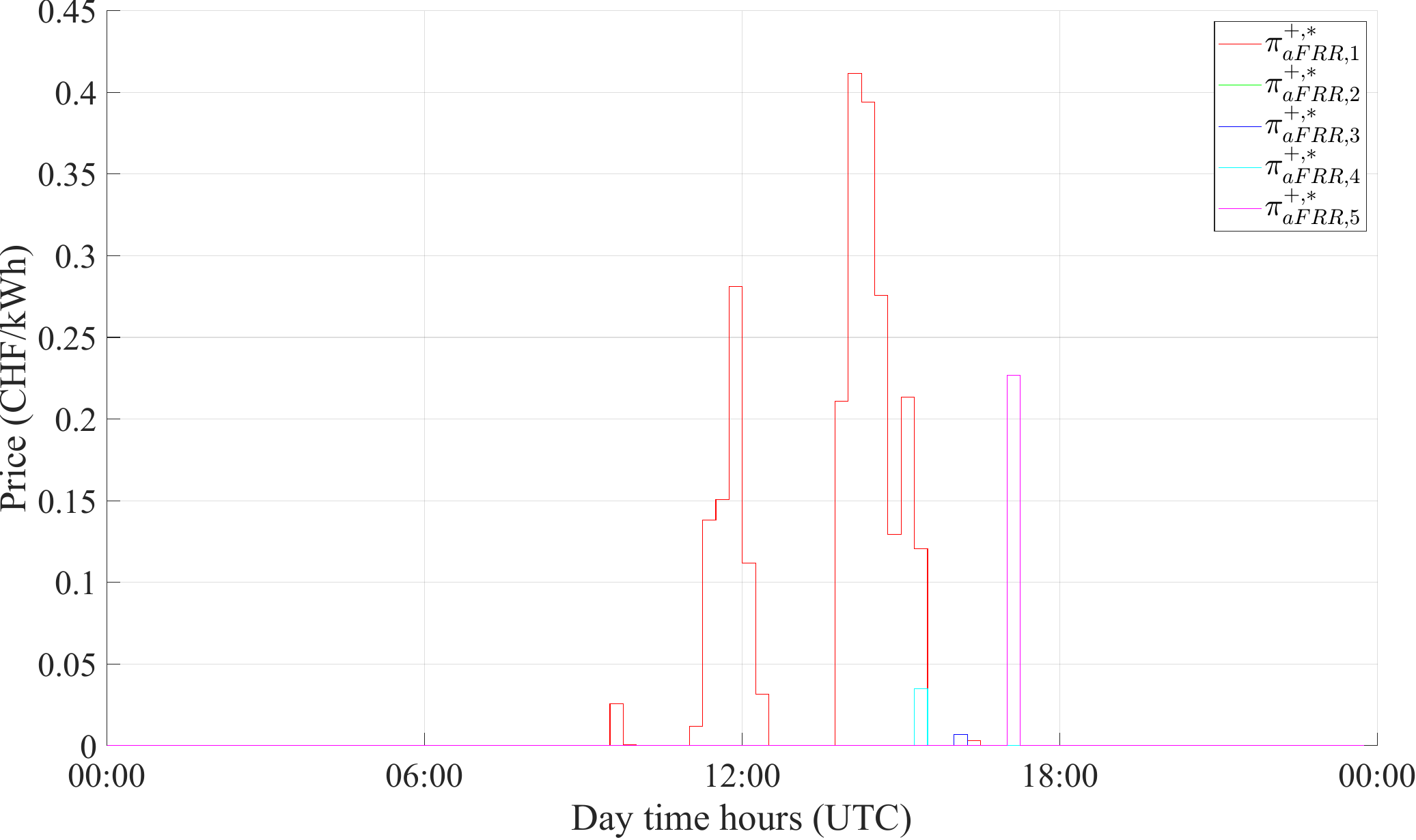}
        \caption{}
        \label{fig.scenarios_pi_aFRR_plus}
    \end{subfigure}

    \begin{subfigure}{\columnwidth}
        \includegraphics[width=\columnwidth]{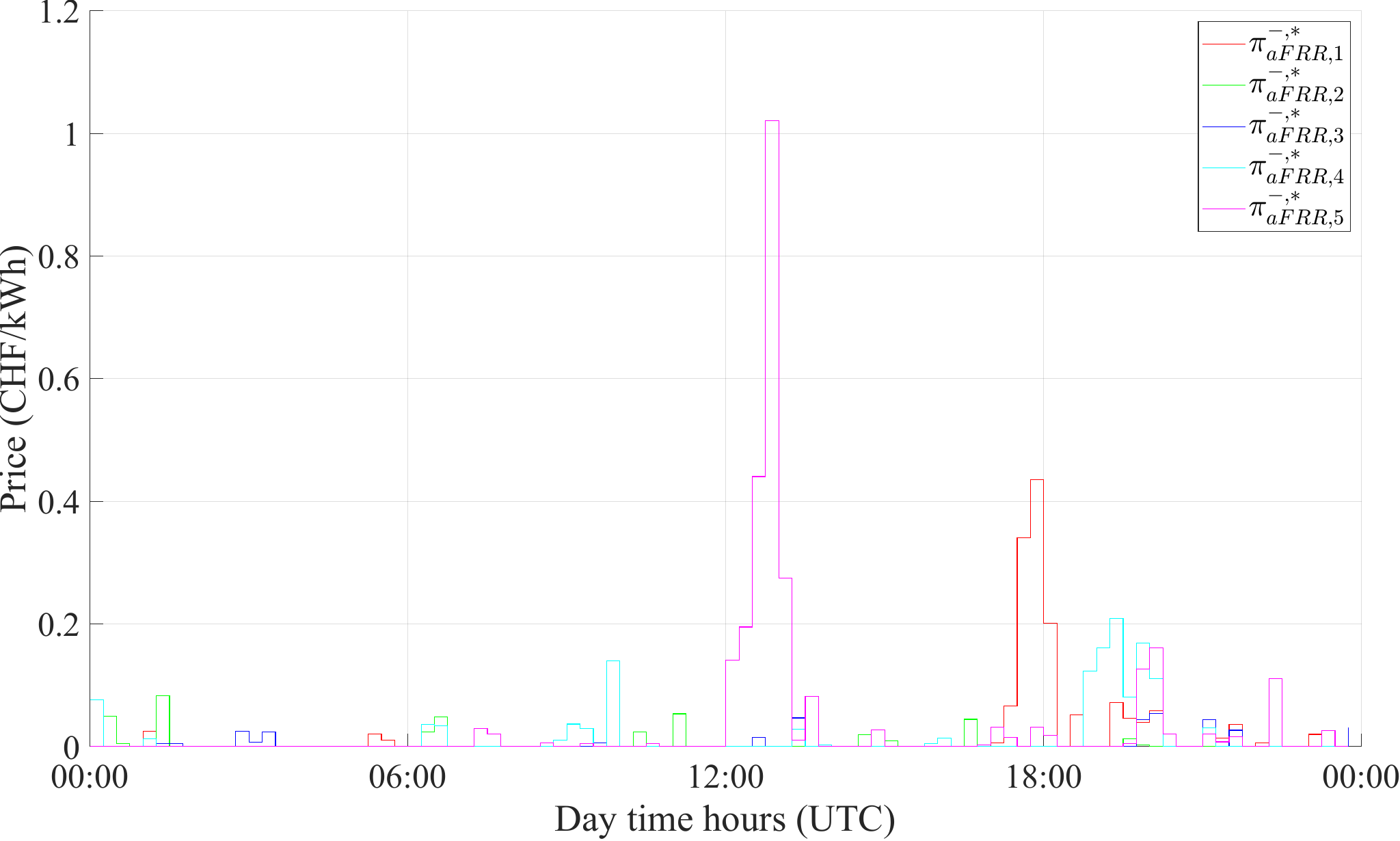}
        \caption{}
        \label{fig.scenarios_pi_aFRR_minus}
    \end{subfigure}

    \caption{aFRR price scenarios for: (a) downward regulation, (b) upward regulation. For improved visualizations, infinite values of $P_{aFRR}^{\pm, *}$ are set to zero.}
    \label{fig.scenarios_pi_aFRR_star}
\end{figure}

Electric energy is billed under a bi-level retail electricity tariff \citep{oiken2024}, with peak and off-peak prices of 0.171~CHF/kWh and 0.162~CHF/kWh, respectively, according to the schedule shown in Fig.~\ref{fig.pi_afrr_star}. In addition, a peak-demand charge of 150~CHF/kW/year is applied, and the feed-in tariff for excess electricity injection is set to 0.0068~CHF/kWh. aFRR price data are obtained from the Swissgrid website \citep{swissgrid2024} and are used to generate scenarios for the day-ahead scheduling stage. The considered scenarios correspond to five consecutive days selected from historical aFRR data to capture representative price variations, including period of which the aFRR price exceeds the retail electricity tariffs. All scenarios are assumed to be equally probable (Fig.~\ref{fig.scenarios_pi_aFRR_star}).

In practice, no contractual agreement exists with the grid operator for the provision of aFRR; therefore, the experiments are performed with the objective of validating the proposed method. The aFRR market prices considered in the day-ahead scheduling correspond to historical market prices. However, for real-time control, since these data are available in real time only to market participants, real-time market prices are synthetically generated by drawing from random distributions, under the operational assumption that regulation requests are issued at a 30-second control resolution and may require either upward or downward regulation, or no activation at all.
This activation structure is consistent with aFRR operational rules in several countries \citep{swissgrid2024_aFRR,elia}. The proposed formulation can also be applied in the context of aggregators, which combine the contributions of multiple small prosumers to provide services to the grid operator.

\subsection{IT setup and actuation layer}
The day-ahead scheduling and MPC controller are executed on a desktop computer equipped with an Intel Core i7 CPU and 16 GB of RAM. The optimization problems are formulated in MATLAB/CVX and solved using Gurobi. The day-ahead scheduling task is triggered daily by the system scheduler. The MPC operates continuously during the day and is briefly paused at midnight to load the newly computed day-ahead schedule, after which it resumes operation using the updated dispatch plan.

Communication with field devices is implemented via standard industrial protocols. In particular, BACnet is used to acquire net-load measurements from the building’s power automation system, while Modbus TCP/IP is employed to communicate with the BESS.
The MPC relies on real-time measurements of the net load and the BESS state. The net load is measured at the building’s main power meter, the BESS active power is obtained directly from the power converter, and the state of energy (SOE) is reported by the battery management system. Photovoltaic (PV) generation data are collected from the inverters; however, these measurements are not used by the real-time MPC, as PV production is inherently reflected in the net-load measurement. Instead, PV data are used in the day-ahead stage for load disaggregation purposes (see section \ref{sec:DA-net:load_forecast}).

\section{Results and Discussion}\label{sec:Results_and_Discussion}
\subsection{Experimental validation}
The objective of the real-time MPC is twofold: (i) to ensure compliance with the dispatch plan by regulating the net power exchanged with the grid, and (ii) to provide aFRR services when economically favorable, while respecting the battery operational constraints. It is worth noting that, since the dispatch plan is designed to ensure the fulfillment of local services (i.e., PV self-consumption and peak shaving), compliance with it implicitly guarantees the provision of these services. The following results illustrate how these objectives are jointly achieved over an entire day of operation.

\subsubsection{Dispatch plan tracking performance}
\begin{figure}[!t]
    \centering
    \begin{subfigure}{\columnwidth}
        \includegraphics[width=\columnwidth]{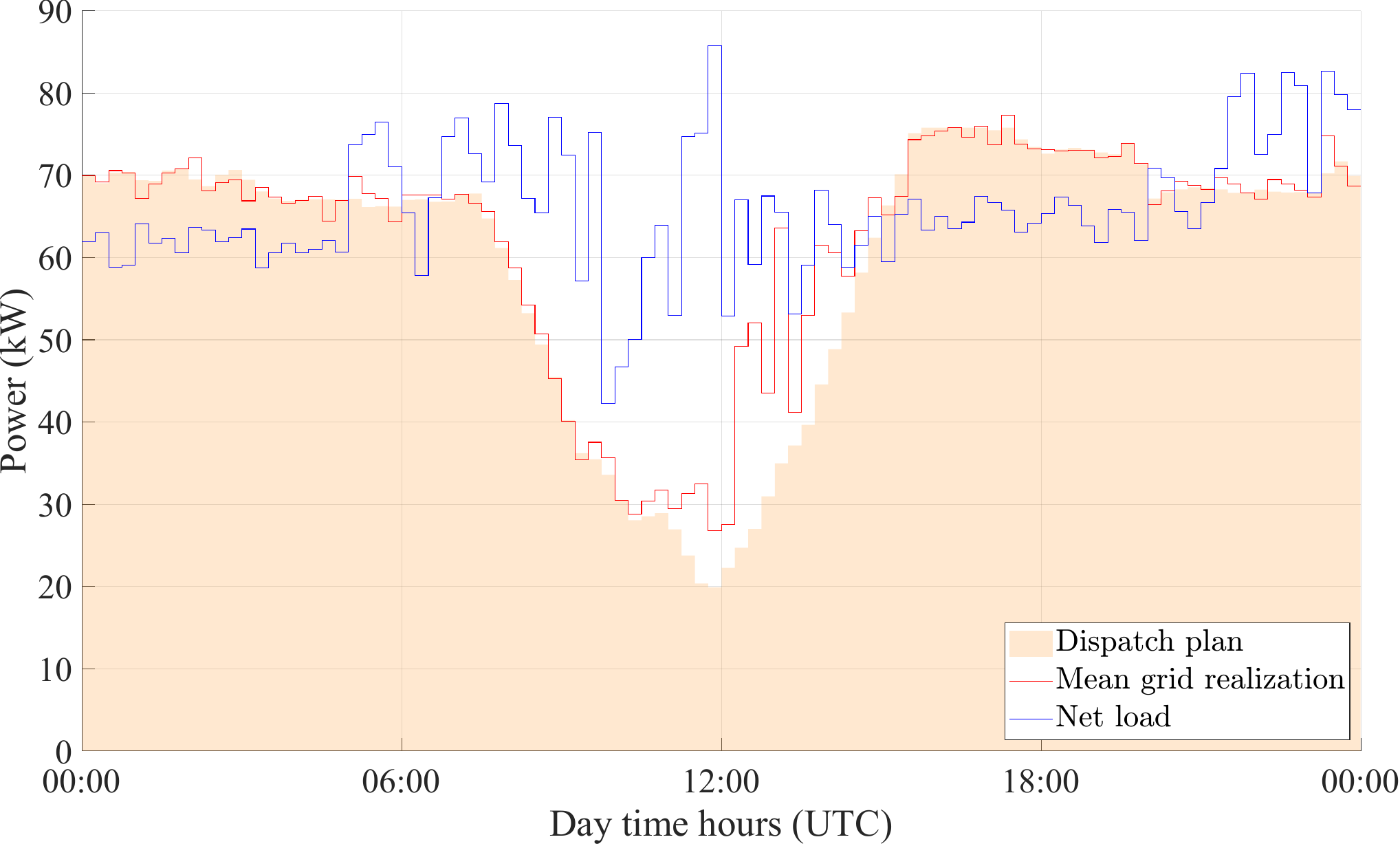}
        \caption{}
        \label{fig.dispatch_plan_tracking}
    \end{subfigure}

    \begin{subfigure}{\columnwidth}
        \includegraphics[width=\columnwidth]{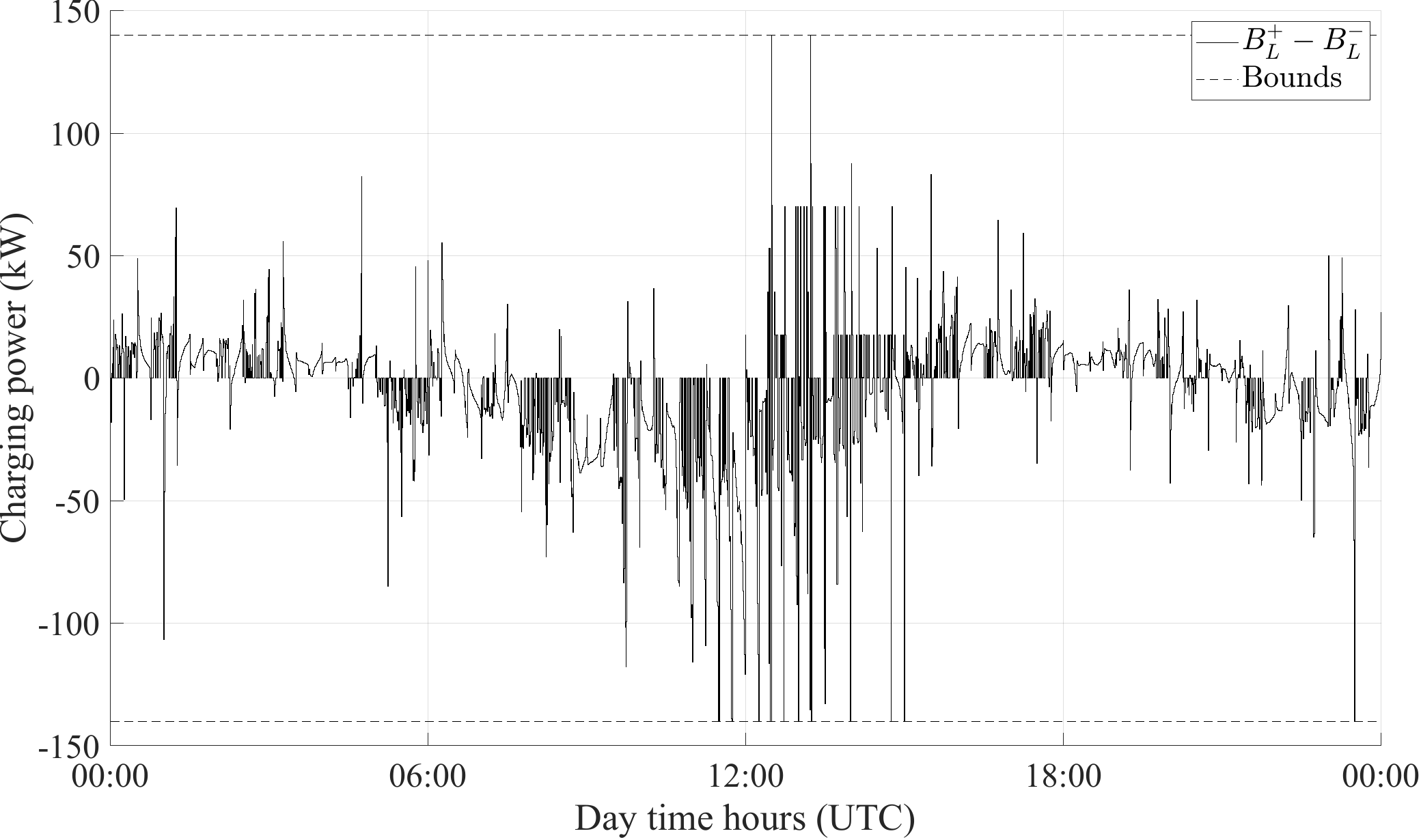}
        \caption{}
        \label{fig.BESS_for_local_services}
    \end{subfigure}

    \begin{subfigure}{\columnwidth}
        \includegraphics[width=\columnwidth]{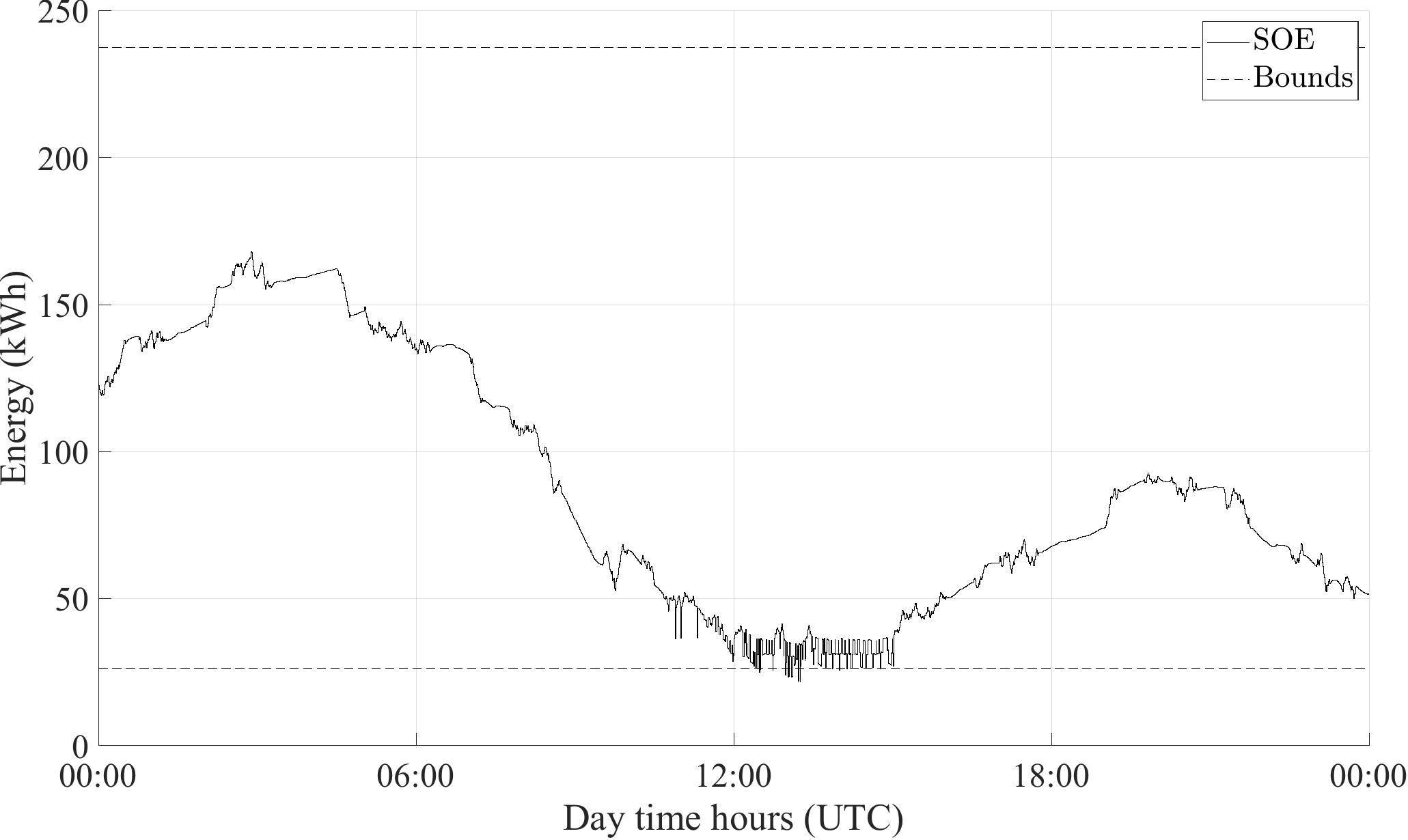}
        \caption{}
        \label{fig.soe}
    \end{subfigure}

    \caption{Real-time results: (a) dispatch plan, actual grid power and net load, averaged over 15-minute intervals, (b) BESS output for local services, (c) BESS's SOE dynamics.}
    \label{fig.local_services}
\end{figure}

Fig.~\ref{fig.local_services} illustrates the performance of the MPC with respect to dispatch plan tracking. Fig.~\ref{fig.dispatch_plan_tracking}) shows the 15-minute averaged dispatch plan of the building complex calculated by the stochastic optimization problem (orange shaded area) and compares it to the average realized power consumption at the grid connection point without aFRR (power flow $P$ represented by the red line) and the average realized power consumption of the buildings (power flow $L$ represented by the blue line). It can be seen that the power at the grid connection point follows the dispatch plan precisely, demonstrating the good tracking performance of the MPC over most period of the day. The corresponding BESS power injections dedicated to dispatch plan tracking are shown in Fig.~\ref{fig.BESS_for_local_services}). The battery alternates between charging and discharging to compensate for deviations between the realized net load and the dispatch plan, thereby smoothing the grid exchange over each dispatch interval.
Fig.~\ref{fig.soe}) depicts the evolution of the BESS state-of-energy (SOE). The SOE remains within the operational bounds for most of the day; however, around midday, the battery reaches its energy limits, resulting in reduced flexibility. This saturation explains the localized deviations from the dispatch plan observed in Fig.~\ref{fig.dispatch_plan_tracking}) during this period. Overall, the coordinated operation enables effective peak shaving, reducing the maximum grid power from 85.74~kW to 77.34~kW, corresponding to a peak reduction of approximately 10\%.

\begin{figure*}[t]%% placement specifier
\centering%% For centre alignment of image.
\includegraphics[width=1\linewidth]{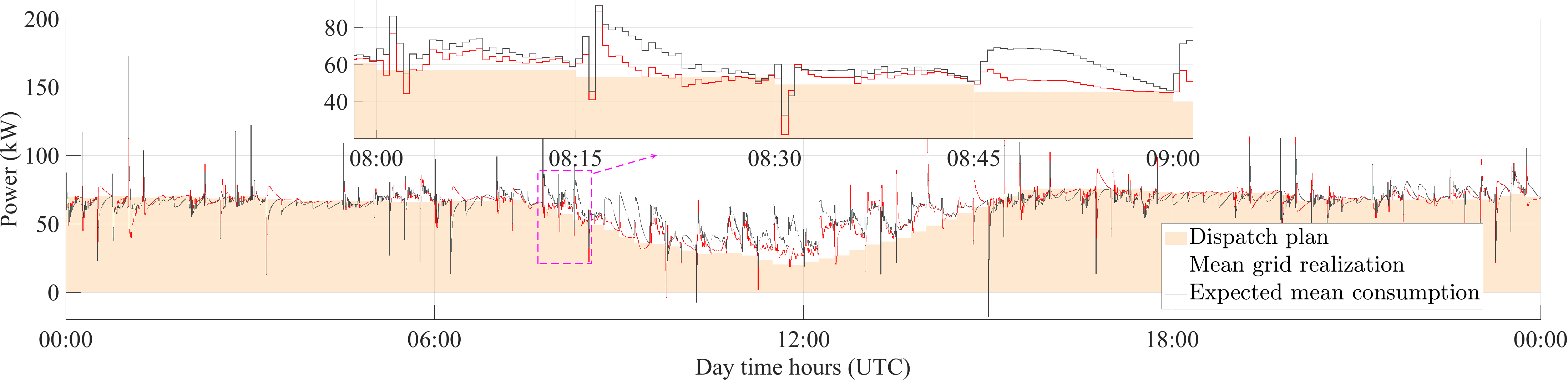}
\caption{Detailed view of the control action of the MPC model.}\label{fig.mpc_details}
\end{figure*}

A detailed overview of MPC operations is depicted in Fig.~\ref{fig.mpc_details}. Within each 15-minute interval, the MPC model controls the BESS to compensate for the mismatch between the net load realizations and the dispatch plan, aiming to achieve a zero-mean error at the end of the interval. It can be seen that, within each interval, the mismatch between grid realizations (red curve) and the dispatch plan decreases over time as a result of the MPC action. Every 30 seconds, the MPC recharges the BESS if the expected average net power blue (black curve)—defined as the average net power over the 15-minute interval combining past grid realizations and predicted future net load—is less than the dispatch plan $\hat{P}(t,k)$, thus increasing consumption to offset this deviation. Conversely, if the expected average net load exceeds the dispatch plan, the BESS is required to discharge.

\subsubsection{aFRR power provision}
\begin{figure}[!t]
    \centering
    \begin{subfigure}{\columnwidth}
        \includegraphics[width=\columnwidth]{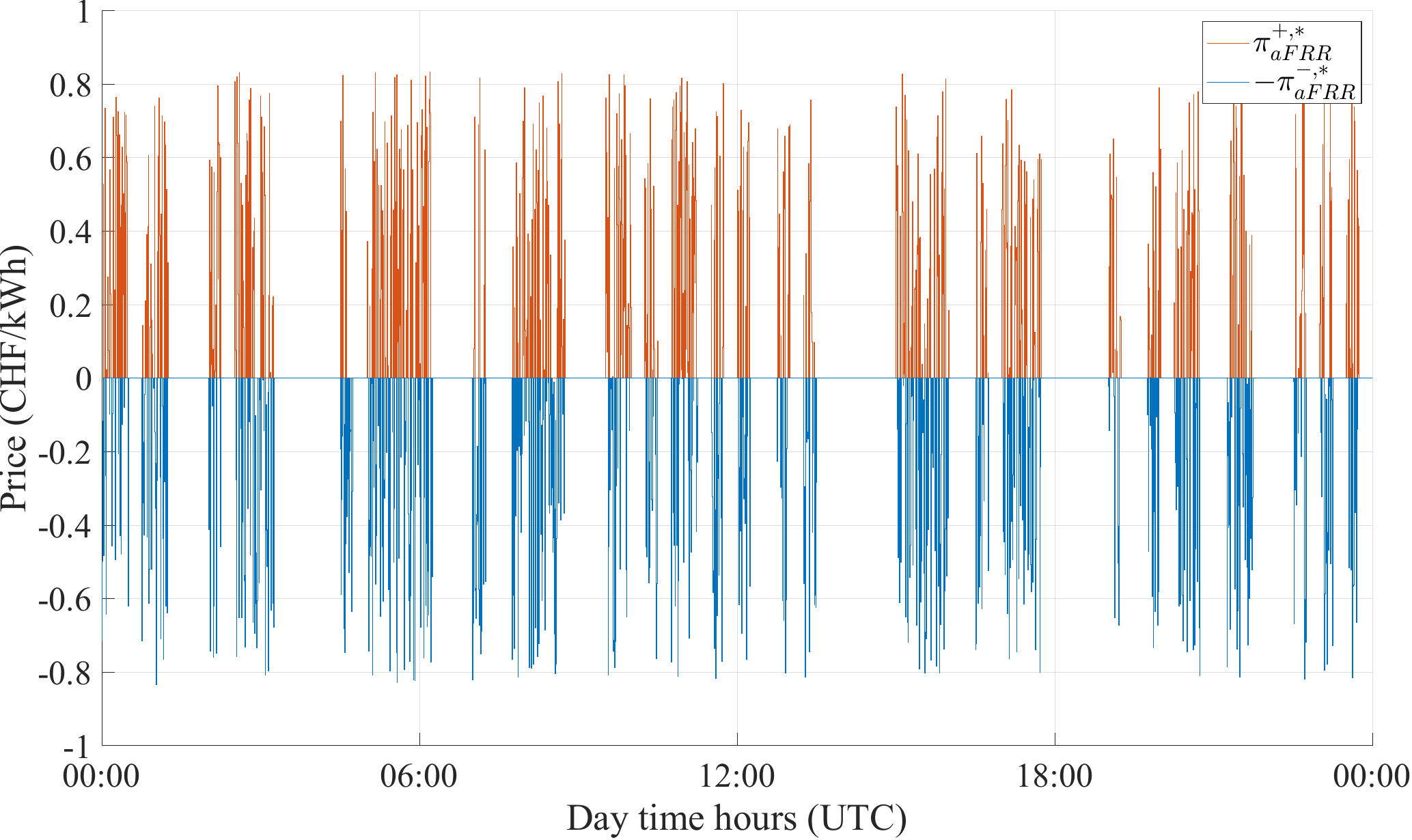}
        \caption{}
        \label{fig.RT_aFRR_prices}
    \end{subfigure}

    \begin{subfigure}{\columnwidth}
        \includegraphics[width=\columnwidth]{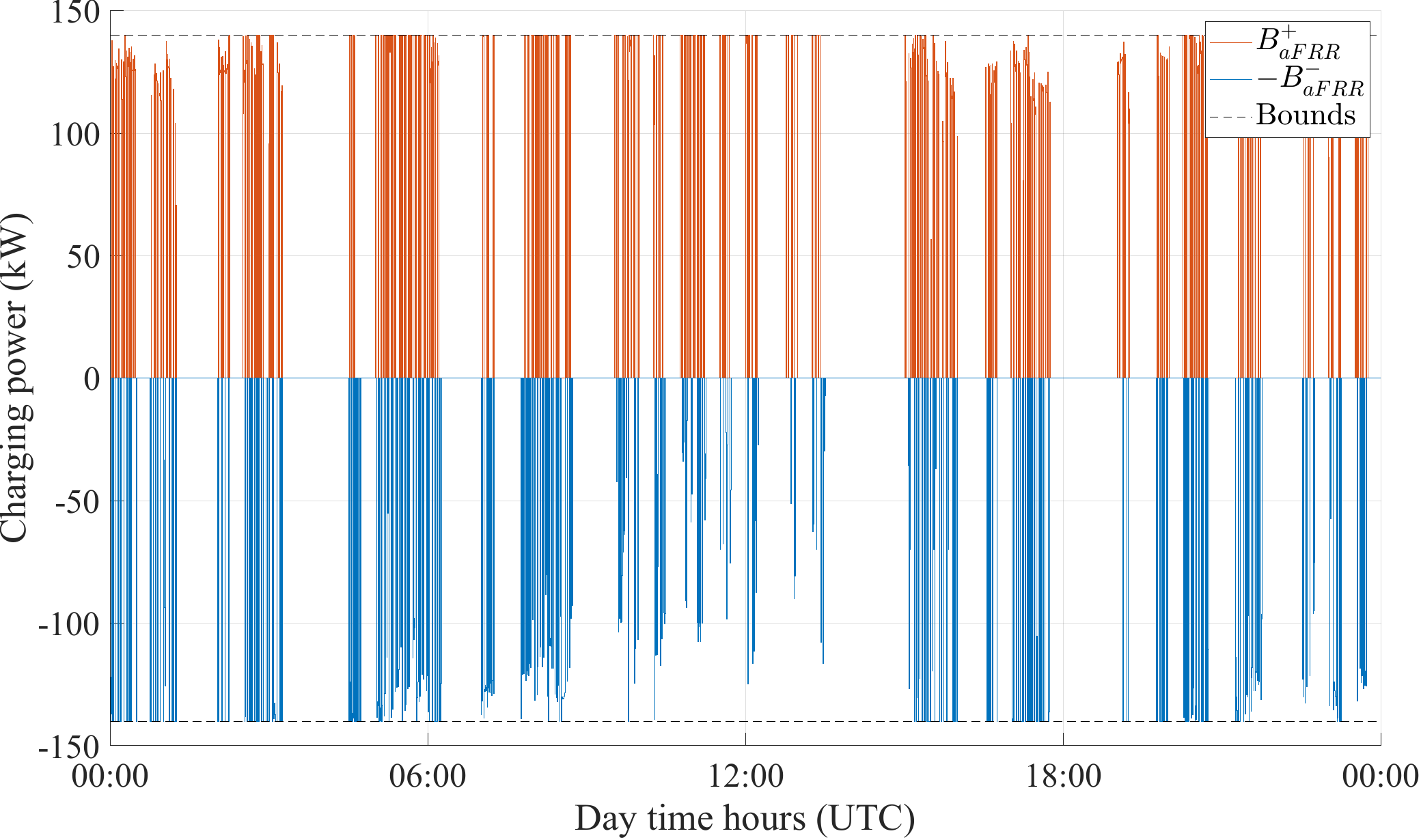}
        \caption{}
        \label{fig.BESS_aFRR_services}
    \end{subfigure}

    \caption{Real-time results: (a) aFRR activation rewards for upward and downward regulation, (b) BESS output for aFRR services. For improved visualizations, infinite values of $P_{aFRR}^{\pm, *}$ are set to zero.}
    \label{fig.aFRR_services}
\end{figure}

The market-driven operation of the BESS is illustrated in Fig.~\ref{fig.aFRR_services}. Fig.~\ref{fig.RT_aFRR_prices}) shows the prices offered by the grid operator for upward and down regulation. Several large values throughout the day denote the remuneration potential from offering flexibility in this market, indicating that sparing some battery power and energy capacity to supply regulating power is worthwhile. Fig.~\ref{fig.BESS_aFRR_services}) shows the BESS contributions to aFRR services. As can be seen, the BESS is taking full advantage of the favorable prices in the regulation markets - while simultaneously providing local services - thus demonstrating that the stochastic optimization problem can effectively spare capacity to achieve this task.

\subsubsection{Coordination of competing services}
\begin{figure}[!t]%% placement specifier
\centering%% For centre alignment of image.
\includegraphics[width=1\linewidth]{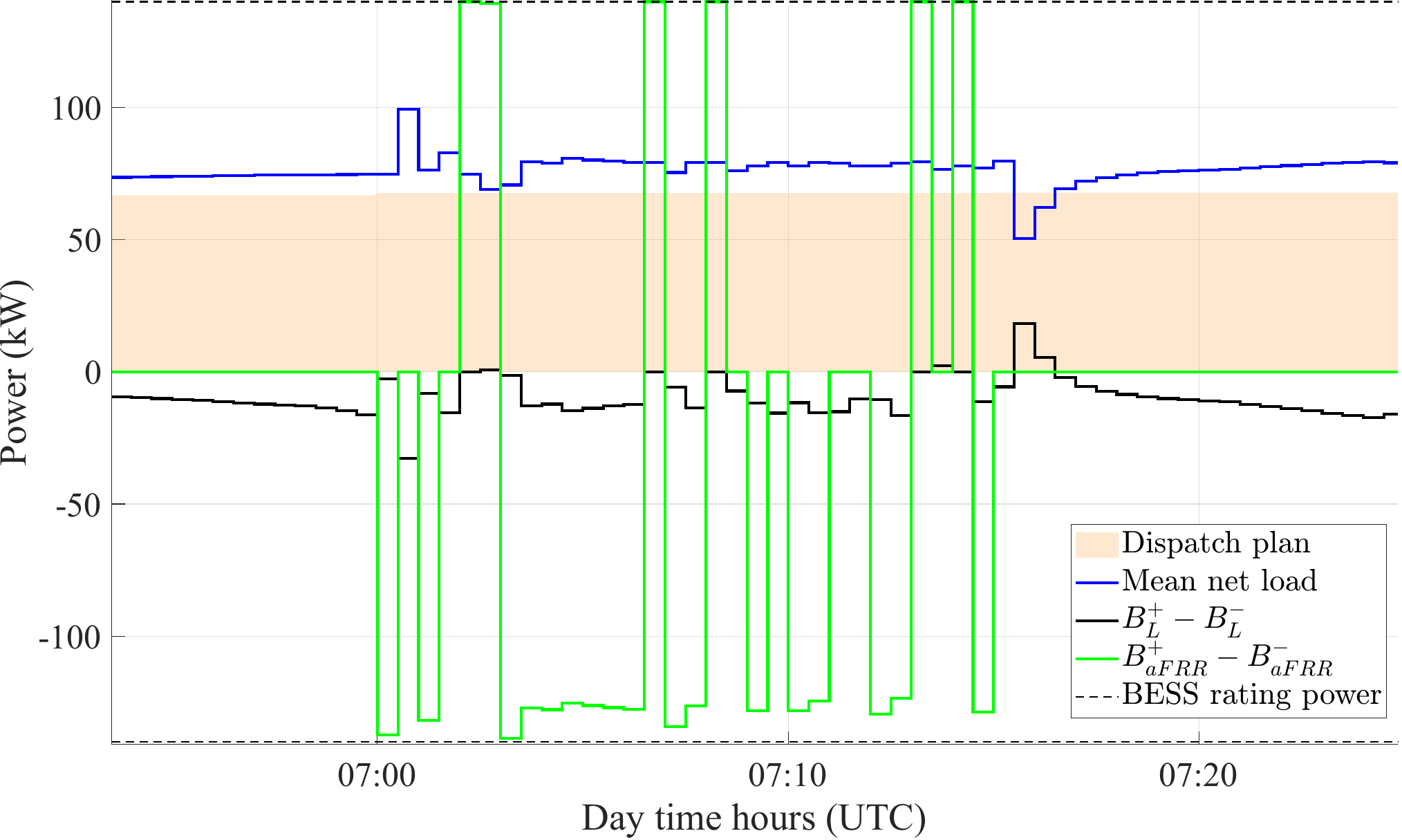}
\caption{Interplay between local and aFRR services within a representative 15 minutes interval.}\label{fig.interplay_local_and_aFRR_services}
\end{figure}

To further clarify the interaction between local services and aFRR provision, an additional zoomed-in view is reported in Fig.~\ref{fig.interplay_local_and_aFRR_services} over a representative 15 minutes interval. The figure shows the dispatch plan in orange, the measured building's net load in blue and the corresponding BESS contributions for both local services in black and aFRR provision in green. The figure highlights how MPC dynamically allocates the BESS power between the two competing services while respecting the BESS's power limits. When aFRR activations occur, the regulating power systematically dominates the local service contribution, reflecting the economic prioritization embedded in the optimization. The remaining capacity is then used for dispatch plan tracking, ensuring compliance with the BESS operational constraints.

\subsection{Economic performance}
The economic performance of the proposed control strategy is evaluated in terms of energy costs and ancillary service revenues. In the absence of the BESS, the total electrical energy cost $C_{\text{Energy}}$ amounts to 264~CHF over the considered day. When the BESS is optimally operated, the energy cost is reduced to 248~CHF, highlighting the benefits of dispatch plan tracking and local flexibility provision. For the considered experiment, participation in the aFRR services yields 471 CHF. This value is obtained using synthetically generated aFRR activation price signals and is therefore not intended to represent typical market conditions. Rather, it serves to illustrate the capability of the proposed framework to coordinate local services and ancillary service provision and to exploit regulation opportunities when favorable price signals are present.

\section{Conclusion}\label{sec:Conclusion}
A scheduling and control framework for a smart building equipped with behind-the-meter PV generation and a BESS is presented, enabling the simultaneous provision of local and grid services. The considered services include electricity bill reduction, achieved through improved PV self-consumption and energy arbitrage using the BESS, peak shaving, and the provision of aFRR to the grid operator or an aggregator. The framework consists of two main stages: day-ahead scheduling and real-time operation. In the former stage, a stochastic MILP is solved to account for uncertainties in upward and downward regulation prices and to determine the BESS charging and discharging behavior for each service. In the latter stage, an MPC computes the battery setpoints at high time resolution in a receding-horizon and periodic fashion, based on updated information on load realizations and regulation prices. The proposed methodology is validated in a real-life experimental setting at the smart buildings of the Energypolis Campus of HES-SO Valais in Sion, Switzerland. Experimental results confirm the ability of the proposed scheduling and control strategy to simultaneously provide local services—namely PV self-consumption, peak shaving, and demand optimization under time-varying electricity tariffs—and balancing power.

\section*{Acknowledgment}
This research is supported by Innosuisse in the context of the flagship project STORE (grant agreement 108.230).

\bibliographystyle{elsarticle-num}
\bibliography{biblio_c}



\appendix

\section{Necessary conditions for convexity}
\label{app:convex_reformulation_cost_function}

The electricity cost in Eq.~\eqref{eq:energy_charge} can be written as
\begin{align}\label{eq:reformulate_non_convex_function}
    C_{\mathrm{Energy}} = \bm{c}^\top [\bm{x}]^+ - \bm{d}^\top [\bm{x}]^-,
\end{align}
where $\bm{c}$ and $\bm{d}$ denote the import and feed-in tariffs, respectively, and $[\cdot]^+$ and $[\cdot]^-$ represent the positive and negative parts of a vector. These operators are nonlinear functions of $\bm{x}$. The objective is to establish the convexity of the above expression with respect to the decision variable $\bm{x}$. Using the identity $\bm{x} = [\bm{x}]^+ - [\bm{x}]^-$, it follows that $[\bm{x}]^- = [\bm{x}]^+ - \bm{x}$. Substituting this into~\eqref{eq:reformulate_non_convex_function} yields
\begin{align}\label{eq:reformulate_non_convex_function_details}
    C_{\mathrm{Energy}} 
    &= \bm{c}^\top [\bm{x}]^+ - \bm{d}^\top \left([\bm{x}]^+ - \bm{x}\right) \\
    &= (\bm{c} - \bm{d})^\top [\bm{x}]^+ + \bm{d}^\top \bm{x}.
\end{align}
The resulting expression is the sum of a nonlinear function of $\bm{x}$ and a linear function of $\bm{x}$. Since $[\bm{x}]^+$ is convex in $\bm{x}$, the overall expression is convex provided that all the elements of the vector $\bm{c} - \bm{d}$ are non-negative.

\section{Linear reformulation of the max operator}
\label{app:linearize_max_operator}

Consider a decision vector $\bm{x} = [x_0, x_1, \dots, x_{T-1}]$. 
The nonlinear operator $\max(\bm{x})$ cannot be directly incorporated into a linear optimization problem. 
To obtain an equivalent linear formulation, an auxiliary scalar decision variable $x_{\max}$ is introduced to represent the maximum value of the elements of $\bm{x}$.

This is achieved by imposing the following constraints:
\begin{align}
    x_{\max} \ge x_t, \quad \forall t = 0, \dots, T-1.
\end{align}

If $x_{\max}$ is included in a minimization objective, it attains the smallest value that satisfies these inequalities. 
Consequently, the constraints are binding at optimality, yielding
\begin{align}
    x_{\max} = \max(\bm{x}).
\end{align}

\section{Linear reformulation of the absolute value}
\label{app:abs_term_approximation}

Given with a scalar variable $y$, its absolute value can be expressed as
\begin{align}\label{eq:abs_term_approximation_def}
    |y| = \max(y, -y).
\end{align}
To represent this relation within a linear programming framework, an auxiliary variable $z$ is introduced, subject to the constraints
\begin{align}\label{eq:abs_term_approximation_constraints}
    z &\ge y, \\
    z &\ge -y.
\end{align}
If $z$ is included in a minimization objective and inequalities are satisfied tightly, it holds that
\begin{align}
    z = |y|.
\end{align}

\end{document}